\documentclass[11pt]{article}

\usepackage[a4paper,margin=1in]{geometry}
\usepackage[T1]{fontenc}
\usepackage[utf8]{inputenc}
\usepackage{lmodern}
\usepackage{microtype}
\usepackage{amsmath,amssymb,bm}
\usepackage{booktabs}
\usepackage[table]{xcolor}
\usepackage{array,multirow,multicol}
\usepackage{graphicx}
\usepackage{caption,subcaption}
\usepackage{float}
\usepackage{algorithm}
\usepackage{algpseudocode}
\usepackage{enumitem}
\usepackage{longtable,tabularx}
\usepackage{tcolorbox}
\usepackage[hidelinks]{hyperref}
\usepackage[nameinlink,capitalise,noabbrev]{cleveref}
\usepackage[square,numbers,sort&compress]{natbib}
\usepackage{setspace}
\usepackage{fancyhdr}
\usepackage{titlesec}
\usepackage{mathpazo}
\usepackage{authblk}
\usepackage{hyperref}

\titleformat{\section}{\large\bfseries}{\thesection.}{0.5em}{}
\titleformat{\subsection}{\normalsize\bfseries}{\thesubsection.}{0.5em}{}
\titleformat{\subsubsection}{\normalsize\itshape}{\thesubsubsection.}{0.5em}{}
\titleformat{\paragraph}[runin]{\normalsize\bfseries}{\theparagraph.}{0.5em}{}

\definecolor{headerblue}{RGB}{220,230,242}
\definecolor{rowgray}{RGB}{245,245,245}
\definecolor{AlgoBlue}{RGB}{220,230,241}
\newcolumntype{L}[1]{>{\raggedright\arraybackslash}p{#1}}
\DeclareCaptionFormat{algblue}{%
  \colorbox{AlgoBlue}{%
    \parbox{\dimexpr\linewidth-2\fboxsep}{%
      \strut\textbf{#1:} \textbf{#3}%
    }%
  }%
}
\AtBeginEnvironment{algorithm}{\footnotesize}
\AtBeginEnvironment{algorithmic}{\footnotesize}
\newcommand{\WrapState}[1]{%
  \Statex \hspace{\algorithmicindent}\parbox[t]{\dimexpr\linewidth-\algorithmicindent}{#1}%
}

\graphicspath{{./}{figures/}{Figures/}{images/}{img/}}
\let\Originalincludegraphics\includegraphics
\renewcommand{\includegraphics}[2][]{%
  \IfFileExists{#2}{\Originalincludegraphics[#1]{#2}}{%
    \fbox{\parbox[c][0.23\textheight][c]{0.78\linewidth}{\centering Missing original figure file: \texttt{#2}\\Place the original image in the project folder to render it.}}%
  }%
}

\newenvironment{keywords}{\par\noindent\textbf{Keywords: }}{\par}

\title{\textbf{A Rolling-Horizon Stochastic Optimization Framework for NBA Franchise Management with Distributionally Robust Risk Constraints}\\[0.3em]
\large A Unified Decision Architecture for the New York Knicks}
\author{Siming Zhang, Zhehui Shen, Shijie Chen, Jian Zhou \\
        \small \texttt{siminchang@shu.edu.cn, 2334952611@shu.edu.cn, q2635836894@shu.edu.cn, zhou\_jian@shu.edu.cn} \\
        \vspace{0.5cm}
        \small Shanghai University}

\date{}

\begin{document}
\maketitle
 
\begin{abstract}
NBA franchise management is not a sequence of independent tasks, but a single dynamic control problem in which roster construction, cash-flow discipline, media strategy, external market shocks, and player-health uncertainty interact over time. Using the New York Knicks as a case study, this paper develops a unified decision architecture for franchise management under competitive, financial, and regulatory constraints. The core layer is formulated as a rolling-horizon stochastic mixed-integer program augmented with distributionally robust optimization and conditional value-at-risk constraints, so that long-run franchise value can be optimized while downside exposure remains explicitly controlled. On top of this core layer, we construct coordinated modules for transaction execution, league-expansion shock transmission, media-rights regime transition, and injury-triggered re-optimization. This integrated design reframes multiple managerial mechanisms inside one research problem: how should an NBA franchise allocate resources and update decisions when performance objectives and commercial objectives are jointly determined under uncertainty? The manuscript is organized around problem formulation, model architecture, empirical validation, robustness analysis, and managerial interpretation.
\end{abstract}


\section{Introduction}

\subsection{Background}
\begin{figure}[htbp]
    \centering
    \includegraphics[width=0.6\linewidth]{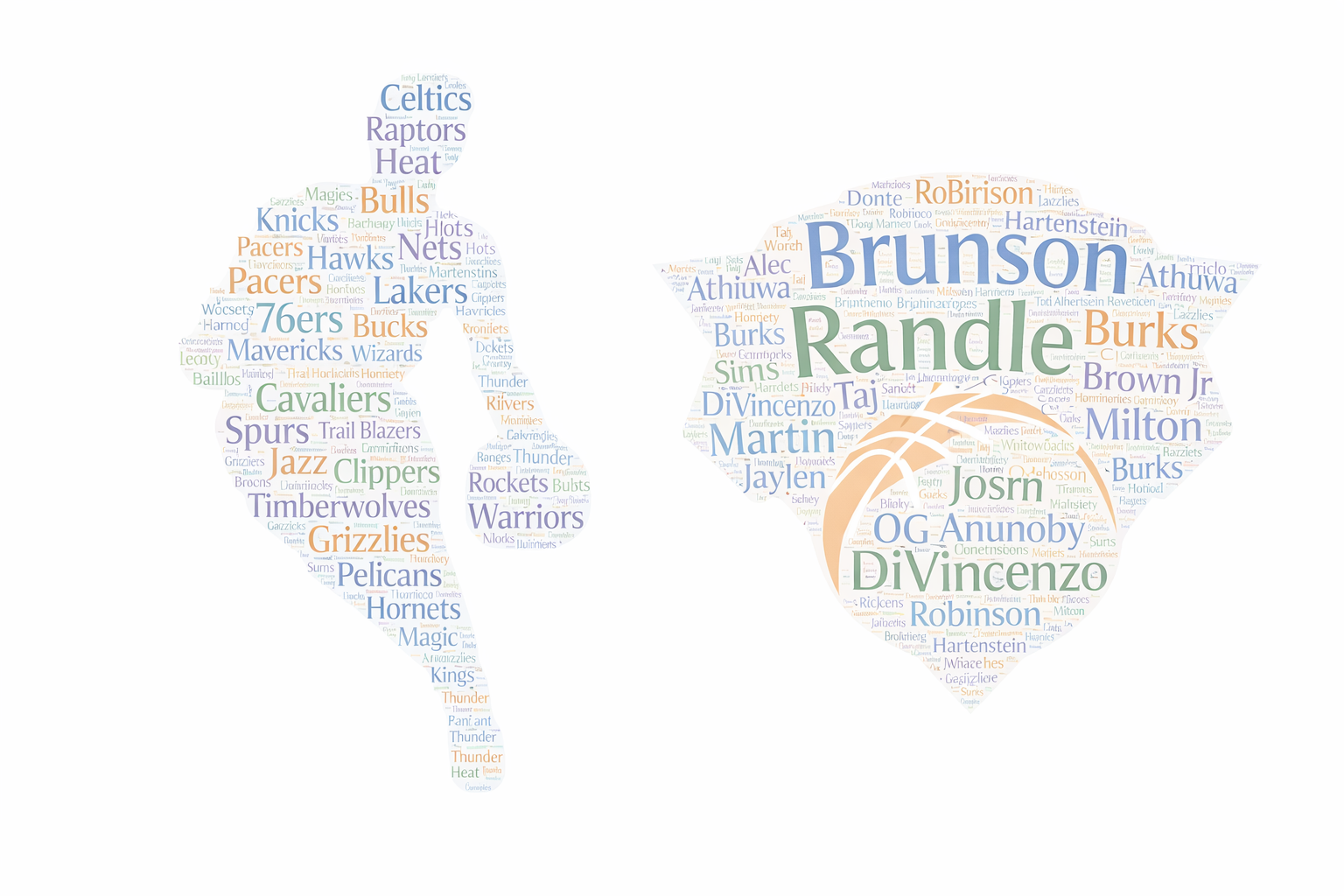}
    \caption{League \& Knicks Identity}
    \label{fig:wordle_overview}
\end{figure}
Professional team sports are best understood as dual-purpose organizations that must jointly manage sporting performance and commercial viability. Classic sports-economics work has shown that league institutions such as player-allocation rules, revenue sharing, and salary restrictions shape franchise incentives rather than merely constraining them ex post \citep{rottenberg1956baseball,fort1995cross}. At the franchise level, empirical valuation evidence further indicates that market size, team performance, and facility effects materially affect team value \citep{alexander2004franchise}. For NBA organizations, this means that roster decisions cannot be evaluated in isolation from local market conditions, brand capital, and the broader league revenue architecture.

This commercial dimension has become even more salient as sports media have shifted from legacy television to hybrid broadcast--streaming ecosystems. The economics of sports rights literature has long emphasized that broadcasting contracts are central to revenue generation and bargaining power in modern leagues \citep{cowie1997rights}. That observation has become more consequential in the NBA after the league's 11-year agreements with Disney, NBCUniversal, and Amazon, which begin with the 2025--26 season and expand distribution across both traditional and direct-to-consumer platforms \citep{nba_media_2024}. Accordingly, any forward-looking franchise model should treat media exposure and platform structure as endogenous components of the decision environment rather than as passive background conditions.

The New York Knicks offer a particularly informative case for this research problem. As a franchise located in the largest U.S. media market and one of the league's most valuable teams, the Knicks exemplify the strategic coexistence of competitive pressure, brand strength, and monetization opportunities \citep{forbes_knicks_2026}. This setting makes the organization a useful empirical anchor for studying how an NBA team should balance roster quality, financial resilience, transaction timing, and shock response inside a single integrated decision system.

\subsection{Research Objective and Central Question}

Rather than treating franchise management as a list of disconnected subproblems, this study formulates a single integrated research question: \emph{how should an NBA franchise jointly optimize roster quality, financial performance, transaction timing, media strategy, and shock response when all decisions are coupled through salary-cap rules, market conditions, and uncertainty?}

Under this perspective, player acquisition, league expansion, media-rights allocation, and injury response are not stand-alone ``tasks.'' They are endogenous modules or exogenous stressors within one franchise-level control system. The objective of the paper is therefore to construct a unified decision architecture that maps data into state variables, state variables into feasible control actions, and control actions into risk-adjusted franchise value.

\subsection{Framework Overview and Contributions}

\begin{figure}[H]
    \centering
    \includegraphics[width=0.7\linewidth]{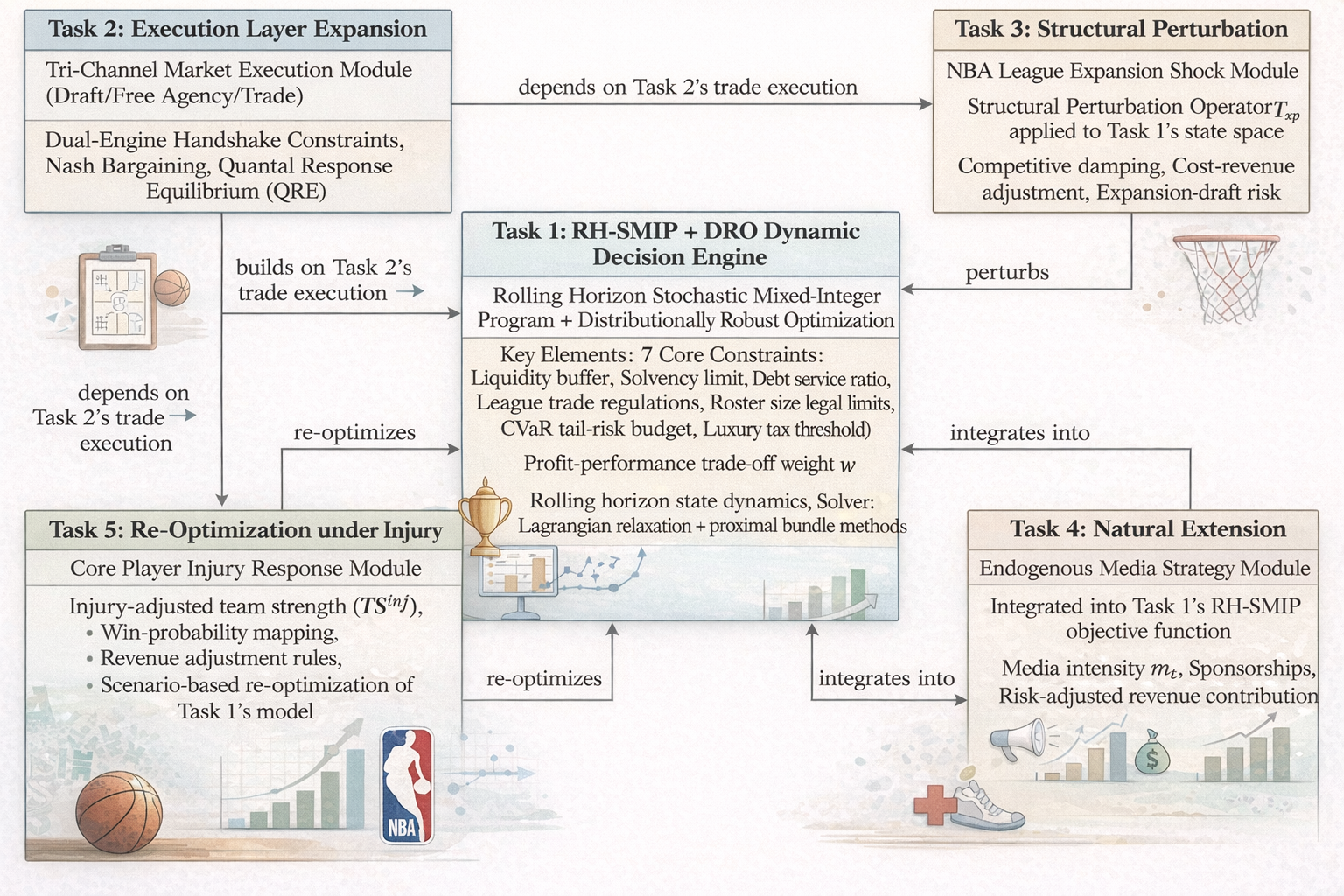}
    \caption{Overview of the NYK-ADMS framework and its coordinated analytical modules.}
    \label{fig:The overview of NYK-ADMS Framework}
\end{figure}

In this research, we develop a unified decision ecosystem, denoted NYK-ADMS, for the New York Knicks. The framework integrates a core franchise optimization layer with four coordinated analytical modules: transaction execution, expansion-shock transmission, media-regime adaptation, and injury-triggered re-optimization. The overall workflow is shown in \Cref{fig:The overview of NYK-ADMS Framework}.

The contributions of the paper can be summarized as follows.
\begin{enumerate}[leftmargin=2em]
    \item We formulate NBA franchise management as a rolling-horizon, risk-constrained optimization problem rather than a sequence of isolated managerial decisions.
    \item We connect roster valuation and market execution through a common surplus-and-risk logic, so that draft, free-agency, and trade decisions inherit the same economic objective as long-horizon planning.
    \item We embed league expansion, media-rights transition, and injury shocks into the same state-space framework, allowing strategic recommendations to be interpreted as responses to structural perturbations rather than as separate case studies.
    \item We translate the resulting optimization outputs into interpretable managerial implications for ownership and franchise-level decision-making.
\end{enumerate}

\section{Related Literature and Research Gap}
The literature most relevant to this study can be organized into three strands.

\subsection{Sports economics, franchise value, and organizational incentives}
Foundational work in sports economics established that professional teams operate inside league-level institutional arrangements that affect both competitive balance and private incentives \citep{rottenberg1956baseball,fort1995cross}. Subsequent empirical work on franchise valuation showed that market size, team success, and facility characteristics all contribute to franchise value formation \citep{alexander2004franchise}. These studies are important because they imply that an NBA front office should not be modeled as a pure win-maximizer: managerial decisions are filtered through local demand, shared league rules, and asset-value considerations.

\subsection{Player valuation, roster construction, and injury analytics.}
A second strand examines how teams evaluate players and assemble rosters. \citet{berri2007wages} argue that conventional basketball narratives frequently misprice individual productivity relative to measurable contribution. In a complementary organizational perspective, \citet{staw1995sunk} show that draft position affects NBA playing time and retention even after controlling for on-court performance, revealing that personnel decisions can embed sunk-cost bias. More recent sports-analytics work has expanded from descriptive metrics to prescriptive roster design. \citet{sarlis2020sports} review modern basketball analytics for player and team evaluation; \citet{muniz2023balanced} develop an optimization-based team-building framework with draft, trade, free-agency, and synergy components; and \citet{ke2024unified} propose a machine-learning framework for NBA and WNBA roster construction under salary information. Injury research has likewise begun to connect performance degradation with economic consequences, rather than treating injuries as purely medical disruptions \citep{sarlis2023injury}.

\subsection{Optimization under uncertainty and the remaining gap.}
A third strand provides the methodological basis for risk-aware decision design. Robust optimization offers tractable protection against adverse realizations when parameters are uncertain \citep{bental2009robust}; conditional value-at-risk provides a coherent and operational tail-risk criterion \citep{rockafellar2000cvar}; and distributionally robust optimization generalizes stochastic programming when the underlying probability law is itself only partially known \citep{delage2010dro}. Although these methods are now standard in operations research, sports applications typically deploy them only for isolated subproblems, such as player evaluation, one-shot roster selection, or injury analysis. What remains comparatively underdeveloped is an integrated franchise-level architecture that simultaneously links roster construction, financial planning, transaction execution, media-regime change, and exogenous shocks within a rolling-horizon state-space model. The present study addresses that gap by synthesizing sports-economics theory, basketball analytics, and risk-aware optimization into a unified framework for NBA franchise management.


\section{Modeling Assumptions}
\begin{itemize}[label=$\heartsuit$]
\item \textbf{Assumption 1:} Coaching ability is independent of player quality.

\textbf{Justification}: A skilled coach can significantly elevate a team's performance. In the NBA, where player quality is uniformly high, there is no scenario where a coach's potential is stifled by subpar players.

\item \textbf{Assumption 2:} Star-driven short-term team revenue is limited to jersey sales and the team's share of individual endorsements.

\textbf{Justification:} Although multiple factors influence short-term team revenue, the primary drivers are jersey sales and the team's share of individual endorsements. Other factors are negligible.

\item \textbf{Assumption 3:} Management decisions are time-consistent across the rolling horizon.

\textbf{Justification:} All re-optimizations follow an unchanged objective and CVaR risk structure, reflecting stable long-term managerial discipline and ensuring dynamic consistency of decisions.

\item \textbf{Assumption 4:} On-court performance affects revenue monotonically with diminishing returns.

\textbf{Justification:} Competitive success always increases revenue but at a decreasing marginal rate, capturing market saturation effects and preventing unrealistic unbounded profit growth.

\end{itemize}


\section{Problem Formulation and Notation}

\vspace*{4pt}

\begingroup
\tiny
\color{black}
\setlength{\parskip}{0pt}
\setlength{\columnsep}{14pt}
\setlength{\premulticols}{6pt}
\setlength{\postmulticols}{0pt}
\setlength{\multicolsep}{3pt}
\renewcommand{\arraystretch}{1.05}
\setlength{\tabcolsep}{3pt}
\rowcolors{2}{rowgray}{white}

\begin{multicols}{2}

\noindent\begin{tabular}{L{0.30\columnwidth} L{0.66\columnwidth}}
\rowcolor{headerblue}\multicolumn{2}{l}{\textbf{Indexing and Sets}}\\
$t$ & Monthly decision epoch index ($t=0,1,\dots,H-1$) \\
$H$ & Planning horizon length (months; e.g., $H=120$) \\
$i$ & Player index \\
$\omega$ & Scenario index \\
$\Omega$ & Set of generated scenarios \\
$I_t$ & Active roster set at month $t$ \\
\end{tabular}

\vspace{0pt}
\noindent\begin{tabular}{L{0.30\columnwidth} L{0.66\columnwidth}}
\rowcolor{headerblue}\multicolumn{2}{l}{\textbf{State Variables (Rolling Horizon)}}\\
$S_t$ & Full system state at month $t$ (roster, finance, exogenous factors) \\
$Cash_t$ & Cash reserve \\
$Debt_t$ & Outstanding debt \\
$DaysStable_t$ & Days of roster stability (time since last roster change) \\
$Streak_t$ & Win/loss streak indicator \\
$Cap_t$ & Salary cap level at month $t$ \\
$Sal_{i,t}$ & Salary of player $i$ at month $t$ \\
$Rem_{i,t}$ & Remaining contract length of player $i$ (months) \\
$Min_{i,t}$ & Minutes played (or expected minutes) of player $i$ \\
$PPS_{i,t}$ & Player Performance Score of player $i$ \\
$Pot_{i,t}$ & Potential score of player $i$ (future upside proxy) \\
$Dur_{i,t}$ & Durability/availability probability of player $i$ \\
$Idur_{i,t}$ & Durability factor used in potential update (scaled availability) \\
\end{tabular}

\vspace{0pt}
\noindent\begin{tabular}{L{0.30\columnwidth} L{0.66\columnwidth}}
\rowcolor{headerblue}\multicolumn{2}{l}{\textbf{Performance Aggregation and Mapping}}\\
$TS_{raw,t}$ & Raw team strength aggregated from player contributions \\
$TS_{adj,t}$ & Adjusted team strength after chemistry/injury/staff effects \\
$C_{chem,t}$ & Chemistry coefficient (streak + stability effect) \\
$I^{DRO}_{inj,t}$ & Worst-case (DRO) injury robustness factor \\
$E_{effect,t}$ & Coaching/staff effect multiplier \\
$Perf_t$ & Team on-court performance metric (e.g., expected win\%) \\
$Wins_t$ & Expected wins or win-rate proxy at month/season scale \\
$f_{perf}(\cdot)$ & Monotone mapping from strength to performance (e.g., logistic) \\
\end{tabular}

\vspace{0pt}
\noindent\begin{tabular}{L{0.30\columnwidth} L{0.66\columnwidth}}
\rowcolor{headerblue}\multicolumn{2}{l}{\textbf{Financial Model}}\\
$Revenue_t$ & Total revenue in month $t$ \\
$Cost_t$ & Total cost in month $t$ \\
$Profit_t$ & Profit in month $t$ ($Revenue_t-Cost_t$) \\
$M_{macro,t}$ & Macro-economic multiplier (city/league conditions) \\
$S_{sat,t}$ & Win-saturation / market saturation factor \\
$K$ & Max saturation level in $S_{sat}(\cdot)$ \\
$r$ & Slope parameter in $S_{sat}(\cdot)$ logistic curve \\
$W_0$ & Half-saturation win level in $S_{sat}(\cdot)$ \\
$TV_t$ & Media/TV baseline revenue component \\
$Gate_t$ & Gate (ticketing) revenue component \\
$Merc_t$ & Merchandise/sponsorship revenue component \\
$R_{star,t}$ & Star-driven incremental brand revenue \\
$B_{i,t}$ & Player $i$ brand score \\
$\gamma$ & Discount factor (monthly) for DCF valuation \\
$Val_{long,H}$ & Terminal/continuation franchise value at horizon end \\
$r_{WACC}$ & Weighted-average cost of capital used in terminal value \\
$g_{growth}$ & Long-run growth rate in terminal value \\
$FCF_H$ & Free cash flow at horizon end (for terminal value) \\
$NPV_H(i)$ & Net present value of player $i$ at horizon end \\
\end{tabular}

\noindent\begin{tabular}{L{0.30\columnwidth} L{0.66\columnwidth}}
\rowcolor{headerblue}\multicolumn{2}{l}{\textbf{Decision Variables (Control Vector)}}\\
$x_t$ & Aggregate decision vector at month $t$ \\
$x_{tick,t}$ & Ticket pricing decisions (base price/discount/premium, etc.) \\
$x_{merc,t}$ & Marketing/sponsorship/merchandising decisions \\
$x_{ven,t}$ & Venue operations decisions (ads, maintenance, concessions) \\
$x_{acq,t}$ & Acquisition module decisions (FA pursuit, rookie dev, etc.) \\
$x_{trade,t}$ & Trade package decisions (players/picks/cash) \\
$x_{cont,t}$ & Contract decisions (extensions, min deals, bonuses) \\
$y^{star}_t$ & Binary indicator: pursue marquee star or not \\
$r^{rookie}_t$ & Rookie development rate / investment intensity \\
$y^{ext}_t$ & Binary indicator: offer extension or not \\
$n^{min}_t$ & Number of minimum contracts signed \\
$r^{bonus}_t$ & Bonus incentive rate \\
\end{tabular}

\vspace{0pt}
\noindent\begin{tabular}{L{0.30\columnwidth} L{0.66\columnwidth}}
\rowcolor{headerblue}\multicolumn{2}{l}{\textbf{Objective and Trade-off}}\\
$w$ & Weight on profit in composite objective ($1-w$ on performance) \\
$J(\cdot)$ & Composite objective value (profit--performance--terminal) \\
\end{tabular}

\vspace{0pt}
\noindent\begin{tabular}{L{0.30\columnwidth} L{0.66\columnwidth}}
\rowcolor{headerblue}\multicolumn{2}{l}{\textbf{DRO (Distributionally Robust Optimization)}}\\
$P$ & Probability distribution over uncertainty \\
$\mathcal P$ & Ambiguity set of distributions (moment-based) \\
$Z_t$ & Uncertain factor vector at month $t$ (macro/injury/etc.) \\
$\hat{\mu}_t$ & Nominal mean of $Z_t$ \\
$\hat{\sigma}_t$ & Nominal standard deviation of $Z_t$ \\
$\rho_{\mu,t}$ & Mean-deviation radius in ambiguity set \\
$\rho_{\sigma,t}$ & Variance-deviation radius in ambiguity set \\
$\mathbb E_{P}[\cdot]$ & Expectation under distribution $P$ \\
\end{tabular}

\vspace{0pt}
\noindent\begin{tabular}{L{0.30\columnwidth} L{0.66\columnwidth}}
\rowcolor{headerblue}\multicolumn{2}{l}{\textbf{CVaR Tail-Risk Budgeting}}\\
$\alpha$ & CVaR confidence level (e.g., $0.95$) \\
$\eta$ & CVaR budget ratio (relative to cap) \\
$\zeta_t$ & VaR threshold (auxiliary variable) at month $t$ \\
$\nu^\omega_t$ & Excess loss beyond $\zeta_t$ in scenario $\omega$ \\
$L^{\omega}_{inj,t}$ & Injury-related loss in scenario $\omega$ at month $t$ \\
$\tau$ & Durability threshold defining ``effectively out'' injury state \\
\end{tabular}

\vspace{0pt}
\noindent\begin{tabular}{L{0.30\columnwidth} L{0.66\columnwidth}}
\rowcolor{headerblue}\multicolumn{2}{l}{\textbf{Core Constraints / Policy Parameters}}\\
$C_{min}$ & Minimum liquidity buffer constraint \\
$D_{max}$ & Maximum allowable debt (solvency limit) \\
$\psi_{DSCR}$ & Minimum debt-service coverage ratio \\
$\theta_{tax}$ & Luxury tax parameters (threshold/rate structure) \\
$\theta_{rules}$ & League rules parameters (trade/roster legality) \\
\end{tabular}

\vspace{0pt}
\noindent\begin{tabular}{L{0.30\columnwidth} L{0.66\columnwidth}}
\rowcolor{headerblue}\multicolumn{2}{l}{\textbf{Algorithmic Symbols (Proximal Bundle + Lagrangian)}}\\
$k$ & Bundle iteration index \\
$K_{max}$ & Max number of bundle iterations per month \\
$\lambda,\mu$ & Dual multipliers (Lagrangian relaxation) \\
$\mathcal B$ & Bundle set of stored subgradients/cuts \\
$u$ & Proximal weight parameter \\
$\epsilon$ & Convergence tolerance \\
$x^{(k)}$ & Primal solution at bundle iteration $k$ \\
$g_k$ & Subgradient at iteration $k$ \\
$e_k$ & Linearization error at iteration $k$ \\
$d_k$ & Dual search step from bundle master QP \\
\end{tabular}

\end{multicols}

\rowcolors{2}{}{ }
\endgroup


\section{Data and Preprocessing}
\subsection{Data Overview}
\begin{figure}[H]
    \centering
    \includegraphics[width=0.9\linewidth]{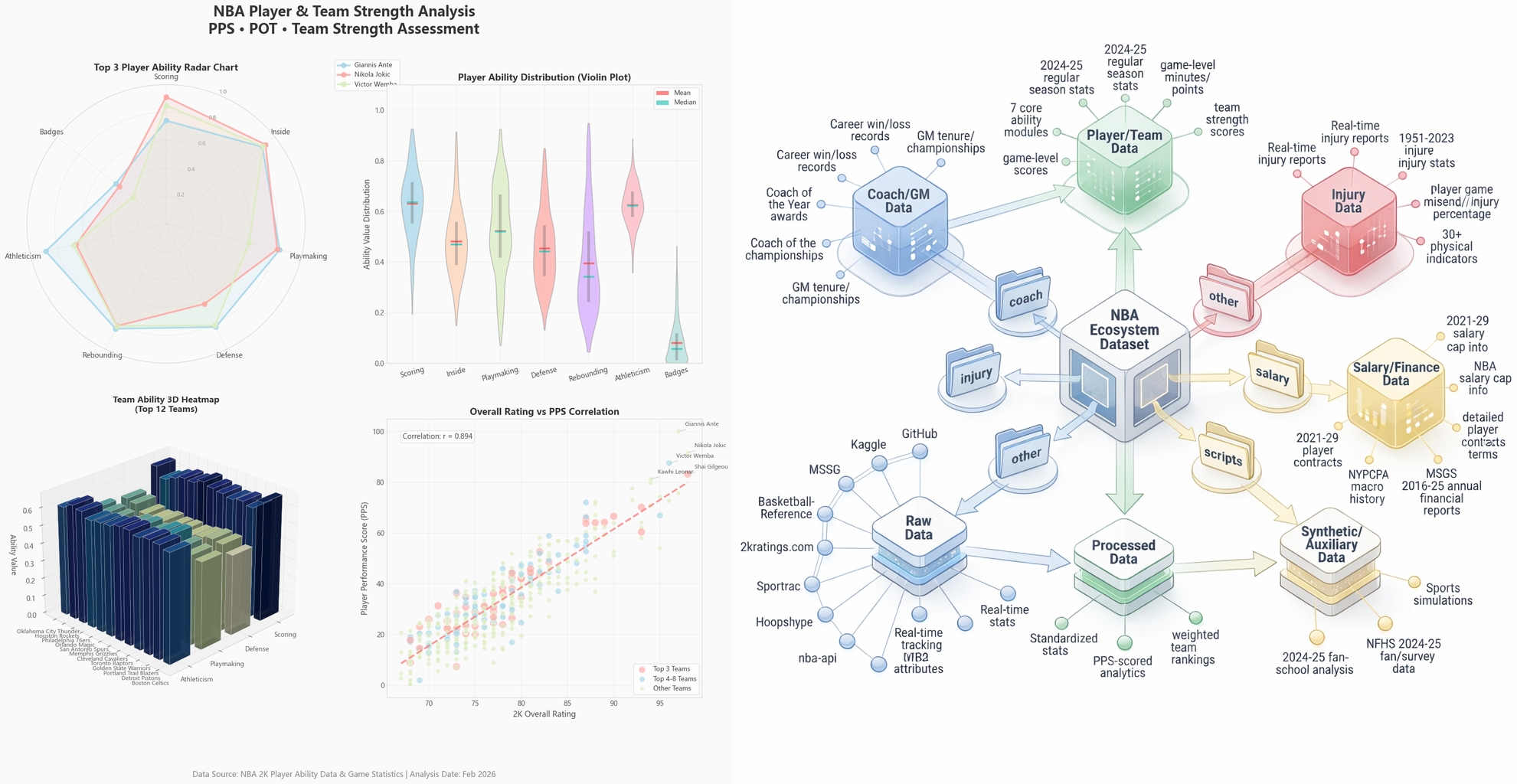}
    \caption{Overview of the dataset}
    \label{fig:The overlook of the data}
\end{figure}
\hspace{1.4em} To support robust modeling of NBA team performance and decision-making, we construct a
multi-source dataset integrating information from multiple organizational levels. All data
are obtained from publicly available and well-documented sources.

The dataset is organized into four hierarchical levels:

\begin{itemize}
  \item \textbf{Player Level:} regular-season statistics (per-game, totals, advanced metrics),
  game-level minutes played, NBA~2K skill attributes, injury history, and multi-year salary
  and contract information.
  
  \item \textbf{Team Level:} seasonal win--loss records, team statistical profiles,
  home--away indicators, and opponent information.
  
  \item \textbf{Management Level:} head coach career records (regular season and playoffs) and general manager tenure and background characteristics.
  
  \item \textbf{Macro Level:} salary cap history, regional consumer price index, and annual financial reports of Madison Square Garden Sports.
\end{itemize}

Primary data sources include Basketball-Reference, nba-api, HoopsHype, Spotrac, NBA~2K ratings,
FRED, Kaggle, and publicly available GitHub repositories.

\subsection{Data Cleaning and Preprocessing}

Data preprocessing is conducted to ensure consistency, interpretability, and reproducibility
prior to model construction.

\subsubsection*{Standardization and Normalization}

\begin{itemize}
  \item \textbf{Entity alignment:} player names and team identifiers are standardized across
  all datasets to resolve formatting differences, abbreviations, and suffixes.
  
  \item \textbf{Numerical normalization:} all monetary variables are converted to numeric
  format; playing-time data are uniformly expressed in minutes; dates are standardized.
\end{itemize}

\subsubsection*{Missing Values and Outliers}

\begin{itemize}
  \item Missing ratio-based statistics are recomputed when feasible; injury availability is
  estimated using recent multi-season participation rates.
  
  \item Extreme or logically inconsistent observations are filtered using domain-based thresholds, while rookies and two-way contract players are retained with indicator flags.
\end{itemize}

\subsubsection*{Cross-Table Integration and Reproducibility}

\begin{itemize}
  \item
  All hyperparameter values are obtained through ridge regression applied to NBA historical data.
  \item Player-level metrics are aggregated to the team level using consistent season and team
  keys. Team strength indicators are constructed as weighted combinations of player
  performance measures.
  
  \item Raw and processed data are stored separately, and all preprocessing steps are
  implemented through documented scripts and notebooks to ensure full reproducibility.
\end{itemize}

\section{Unified Decision Architecture}
\begin{figure}[H]
    \centering
    \includegraphics[width=0.8\linewidth]{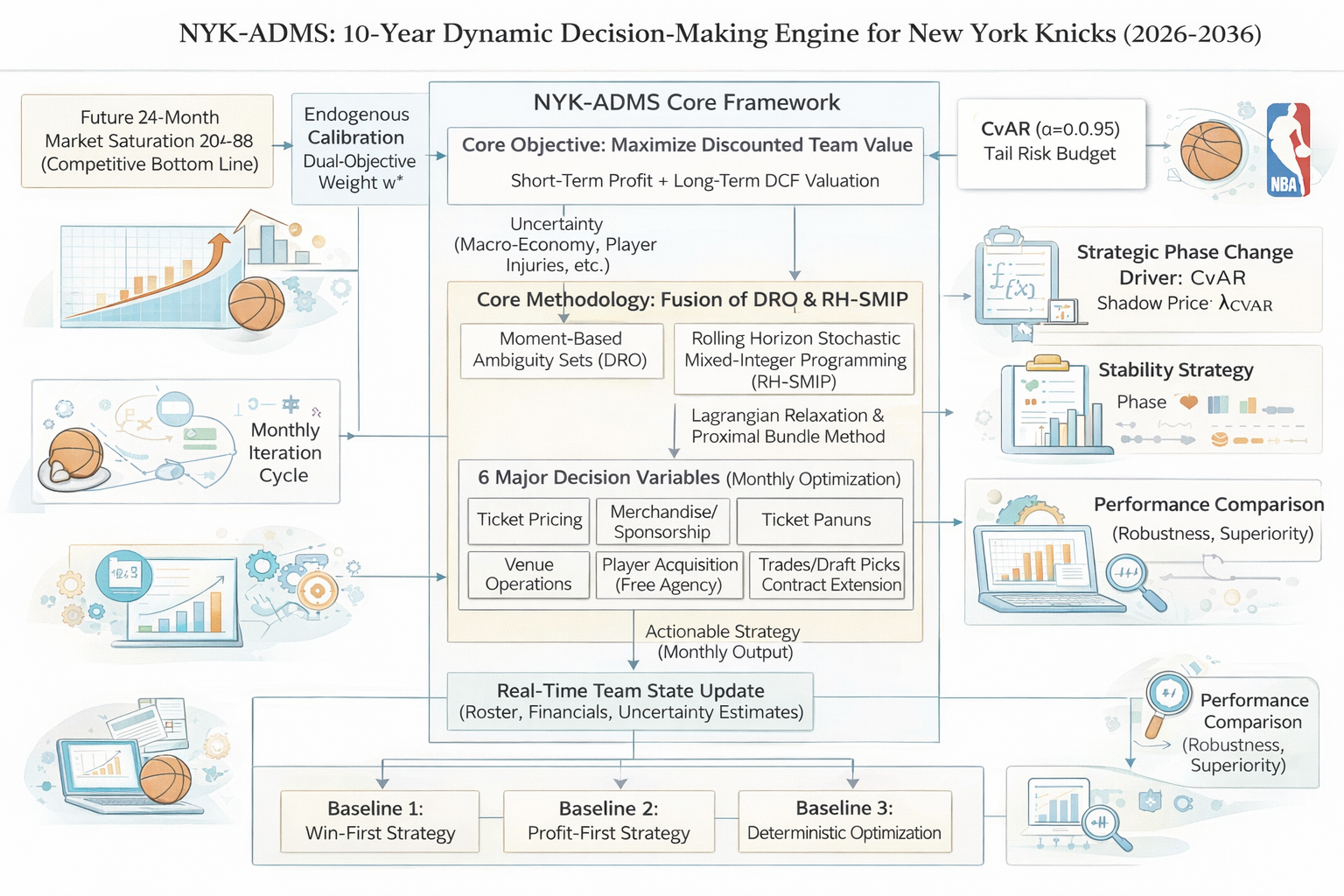}
    \caption{NYK-ADMS: 10-Year Dynamic Decision-Making Engine for New York Knicks}
    \label{fig:NYK-ADMS: 10-Year Dynamic Decision-Making Engine for New York Knicks (2026-2036)}
\end{figure}

 \hspace{1.4em} We formulate a robust rolling-horizon stochastic mixed-integer program (RH-SMIP) as the core decision engine, integrating a distributionally robust optimization (DRO) framework with an explicit CVaR tail-risk budget. This model optimizes the NBA franchise's long-term strategy under uncertainty in performance and economic conditions. A CVaR constraint limits downside risk from injuries, while the objective maximizes expected franchise value. We solve the large-scale robust RH-SMIP via Lagrangian relaxation and a proximal bundle method. The engine operates in a rolling manner: at each monthly decision step it ingests the current state $S_t$, solves the RH-SMIP for the remaining horizon, implements the first-month decisions $x_t^*$, and then updates the state for the next month under realized outcomes.

\subsection{Rolling Horizon Framework and State Dynamics}

\hspace{1.4em} Decision stages are indexed by $t=0,1,\ldots,H-1$, representing months over a 120-month planning horizon (2026--2036). At each month $t$, the state $S_t$ contains all information needed for optimization: roster status, financial metrics, and relevant exogenous factors. Specifically, for each player $i\in I_t$ on the roster, $S_t$ tracks their performance score $\text{PPS}_{i,t}$, potential $\text{Pot}_{i,t}$, durability $\text{Dur}_{i,t}$ (health probability), salary $\text{Sal}_{i,t}$, and remaining contract length $\text{Rem}_{i,t}$. Team-level components include cash reserves $\text{Cash}_t$, debt $\text{Debt}_t$, win streak $\text{Streak}_t$, roster stability $\text{DaysStable}_t$, salary cap $\text{Cap}_t$, and other external conditions.

\begingroup
\small
\paragraph*{Complete state vector $S_t$}
For completeness, we formally define the system state at month $t$ as 
\[ 
\begin{aligned}
S_t 
= 
\Big(
&\underbrace{ 
\{\text{PPS}_{i,t},\ \text{Pot}_{i,t},\ \text{Dur}_{i,t},\ \text{Sal}_{i,t},\ \text{Rem}_{i,t}\}_{i\in I_t} 
}_{\text{roster / contract}},\; 
\underbrace{ 
\text{Cash}_t,\ \text{Debt}_t 
}_{\text{balance sheet}}, 
\\ 
&\underbrace{ 
\text{Streak}_t,\ \text{DaysStable}_t,\ \text{WinRate}_{\text{recent},t} 
}_{\text{chemistry / coaching}},\; 
\underbrace{ 
\text{Cap}_t,\ \mathcal{E}_t 
}_{\text{league / macro}} 
\Big)\,,
\end{aligned}
\] 
with groups corresponding to player roster/contract attributes, team balance-sheet variables, team chemistry/coaching metrics, and league-level or macroeconomic factors, respectively. All symbols introduced in the subsequent sections are either state components, user-specified parameters, or quantities induced by the core franchise optimization layer.
\endgroup

\hspace{1.4em} At time $t$, given the current state $S_t$, the model chooses a decision vector $x_t$ (e.g. pricing, acquisitions, contracts, operations). Then $S_{t+1}$ is realized through deterministic state transitions and stochastic shocks (performance fluctuations, injuries). The main state update equations each month are: 
\[ 
\left\{ 
\begin{aligned} 
\text{Cash}_{t+1} &= \text{Cash}_t + \text{Profit}_t - \text{DebtService}_t,\\ 
\text{Debt}_{t+1} &= \text{Debt}_t + \Delta \text{Debt}_t,\\ 
\text{DaysStable}_{t+1} &= 
\begin{cases} 
\text{DaysStable}_t + 30, & \text{if roster unchanged during month $t$,}\\ 
0, & \text{otherwise,} 
\end{cases}\\ 
\text{Streak}_{t+1} &= \text{UpdateStreak}(\text{Streak}_t,\, \text{Wins}_t),\\ 
\text{PPS}_{i,t+1} &= \text{PPS}_{i,t} + \Delta \text{PPS}_{i,t},\\ 
\text{Pot}_{i,t+1} &= \text{PotUpdate}(\text{PPS}_{i,t+1},\, \lambda_i,\, \Delta_i,\, I_{dur,i,t+1}),\\ 
\text{Dur}_{i,t+1} &= \text{DurUpdate}(\text{Dur}_{i,t},\, \text{InjuryEvent}_{i,t}),\\ 
\text{Rem}_{i,t+1} &= \max\{\text{Rem}_{i,t}-1,\,0\}, \qquad \forall i\in I_t~. 
\end{aligned} 
\right. 
\] 

Here $\Delta \text{PPS}_{i,t}$ and $\text{InjuryEvent}_{i,t}$ are scenario-driven shocks, and $\lambda_i, \Delta_i$ are player-specific constants. These modular update rules align with the distributionally robust scenario generation (introduced later). After each month, the state is updated via these equations and the planning horizon rolls forward.

\subsection{Performance Metrics and Team Dynamics}

\hspace{1.4em} We quantify on-court impact with a composite \textbf{Player Performance Score (PPS)} for each player $i$ at time $t$, defined as a weighted sum of performance stats:
\[ PPS_{i,t} = w_O O_{i,t} + w_I I_{i,t} + w_P P_{i,t} + w_D D_{i,t} + w_R R_{i,t} + w_A A_{i,t} + w_B B_{i,t}, \] 
where $O_{i,t}, I_{i,t}, \ldots, B_{i,t}$ are the player's offensive, defensive, and other stat categories, and $w_\cdot$ are predetermined weights. We also define a \textbf{potential score (Pot)} as a proxy for player $i$'s future upside:
\[ Pot_{i,t} = 100\Big((1-\lambda_i)\,PPS_{i,t} \;+\; \lambda_i\, \frac{\min(PPS_{i,t}+\Delta_i,\,99)}{99}\, I_{dur,i,t}\Big). \] 
Here $\lambda_i \in [0,1]$ weights latent potential vs. current performance, $\Delta_i$ is the player's upside (gap to the benchmark rating of 99), and $I_{dur,i,t}\in[0,1]$ is a durability factor (e.g. $0$ if a serious injury at $t$ curtails development, $1$ if fully healthy). Thus, $Pot_{i,t}$ is a 0--100 scaled $PPS$ that is boosted for high-upside, healthy players. 

\hspace{1.4em} At the team level, individual contributions aggregate into a \textbf{raw team strength}: 
\[ TS_{raw,t} = \sum_{i\in I_t} \frac{E[\text{Min}_{i,t}]}{48}\;PPS_{i,t}, \] 
where $E[\text{Min}_{i,t}]$ is the expected fraction of game minutes (out of 48) that player $i$ plays. This raw strength is then multiplied by several factors: a \textbf{chemistry coefficient} $C_{chem,t} = 1 + \gamma_1\, Streak_t + \gamma_2 \ln(DaysStable_t + e)$ to capture momentum and stability (with small parameters $\gamma_1,\gamma_2$); a \textbf{robust-injury factor} $I_{inj,t}^{DRO} \in (0,1]$ to account for worst-case injury impacts (e.g. $I_{inj,t}^{DRO} = \min_{P\in \mathcal{P}_{inj,t}} E_P[e^{-\kappa L_{inj,t}}]$ for an injury loss metric $L_{inj,t}$); and a \textbf{coaching/staff effect} $E_{effect,t} = \textit{CoachScore}\cdot WinRate_{\text{recent},t} + \alpha_s \ln(E_{staff,t})$ which grows with recent performance and training investment $E_{staff,t}$. The \textbf{adjusted team strength} is 
\[ TS_{adj,t} = TS_{raw,t} \cdot C_{chem,t} \cdot I_{inj,t}^{DRO} \cdot E_{effect,t}, \] 
which drives the overall performance metric $Perf_t = f_{perf}(TS_{adj,t})$. We calibrate $f_{perf}$ so that a given $TS_{adj,t}$ corresponds to a target win rate (e.g. ensuring a minimum strength yields a 48\% win rate floor). In practice, a linear approximation $Wins_t \approx a_0 + a_1\,TS_{adj,t}$ can be used. Thus, $Perf_t$ is a normalized measure of team success (expected win percentage) based on player performance and the above factors.

\subsection{Financial Model: Revenue, Cost, and Terminal Value}

\hspace{1.4em} Each month, the franchise's \textbf{profit} is $Profit_t = Revenue_t - Cost_t$. We decompose revenue into components. A \textbf{macro-economic multiplier} $M_{\text{macro},t}$ captures external market conditions (e.g. local income and competition from other sports). For example: 
\[ M_{\text{macro},t} = \frac{DispIncome_{NY,t}}{BaselineDI} \cdot (1 - Unemployment_t)\cdot (1 - \delta\, \textit{AttnAltSports}_t), \] 
where $\textit{AttnAltSports}_t$ measures local attention on alternative sports and $\delta$ is a sensitivity parameter. The team's success drives demand through a \textbf{wins saturation factor} $S_{sat,t} = S_{sat}(Wins_t)$ modeled by a logistic curve (to reflect ticket market saturation). For instance: 
\[ S_{sat}(Wins) = \frac{K}{\,1 + \exp[-r\,(Wins - W_0)]\,}, \] 
with $K$ the market ceiling, $r$ the slope, and $W_0$ the half-saturation point. 

Given $S_{sat,t}$, we model \textbf{gate revenue} and \textbf{merchandising/sponsorship revenue} as functions of pricing decisions and saturation: $Gate_t = f_{gate}(x_{tick,t}, S_{sat,t})$ and $Merc_t = f_{merc}(x_{merc,t}, S_{sat,t})$, where $x_{tick,t}$ and $x_{merc,t}$ are decision variables for ticket pricing and merchandising/sponsorship. We also include baseline media/TV revenue $TV_t$. Additionally, star players contribute \textbf{brand revenue}: 
\[ R_{star,t} = \sum_{i\in I_t} \beta^* \frac{B_{i,t}}{B_{ref}}, \] 
where $B_{i,t}$ is player $i$'s brand score (with $\beta^*$ and $B_{ref}$ as scaling constants). Thus, total revenue is 
\[ Revenue_t = M_{\text{macro},t}\, S_{sat,t}\,\big( TV_t + Gate_t + Merc_t \big) + R_{star,t}\,. \] 

On the cost side, the model includes player salaries (and luxury taxes), operations, and staff expenses. The \textbf{player salary cost} is 
\[ Cost_{sal,t} = \sum_{i\in I_t} Sal_{i,t} \;+\; \mathrm{LuxuryTax}\Big(\sum_{i\in I_t} Sal_{i,t};\, \theta_{tax}\Big), \] 
where $\theta_{tax}$ defines the luxury tax thresholds and rates. \textbf{Operational costs} include fixed overhead $C_{fixed,t}$ and arena maintenance $b_{maint,t}$ (a decision in $x_{ven,t}$): thus $Cost_{ops,t} = C_{fixed,t} + b_{maint,t}$. \textbf{Staff costs} are given by the coaching/training budget: $Cost_{staff,t} = E_{staff,t}$. Therefore, total cost is $Cost_t = Cost_{sal,t} + Cost_{ops,t} + Cost_{staff,t}$. The monthly profits feed into a discounted cash flow (DCF) valuation of the franchise.

We also include a \textbf{terminal value} at the horizon $t=H$ to capture value beyond the planning window. For example: 
\[ Val_{long,H} = \gamma^H \frac{E[FCF_H]}{\,r_{WACC} - g_{growth}\,} \;+\; \sum_{i\in I_H} NPV_H(i), \] 
where $\gamma$ is the monthly discount factor, $r_{WACC}$ the weighted average cost of capital, $g_{growth}$ the long-term growth rate, $E[FCF_H]$ the expected free cash flow in year $H$, and $NPV_H(i)$ the net present value of any player (or draft pick) $i$ still under contract at $t=H$ (via a scouting model or input). This $Val_{long,H}$ represents the continuation value of the franchise beyond 2036. 

\subsection{RH-SMIP Formulation (DRO Objective and Constraints)}
\begin{figure}[htbp]
    \centering
    \includegraphics[width=0.7\linewidth]{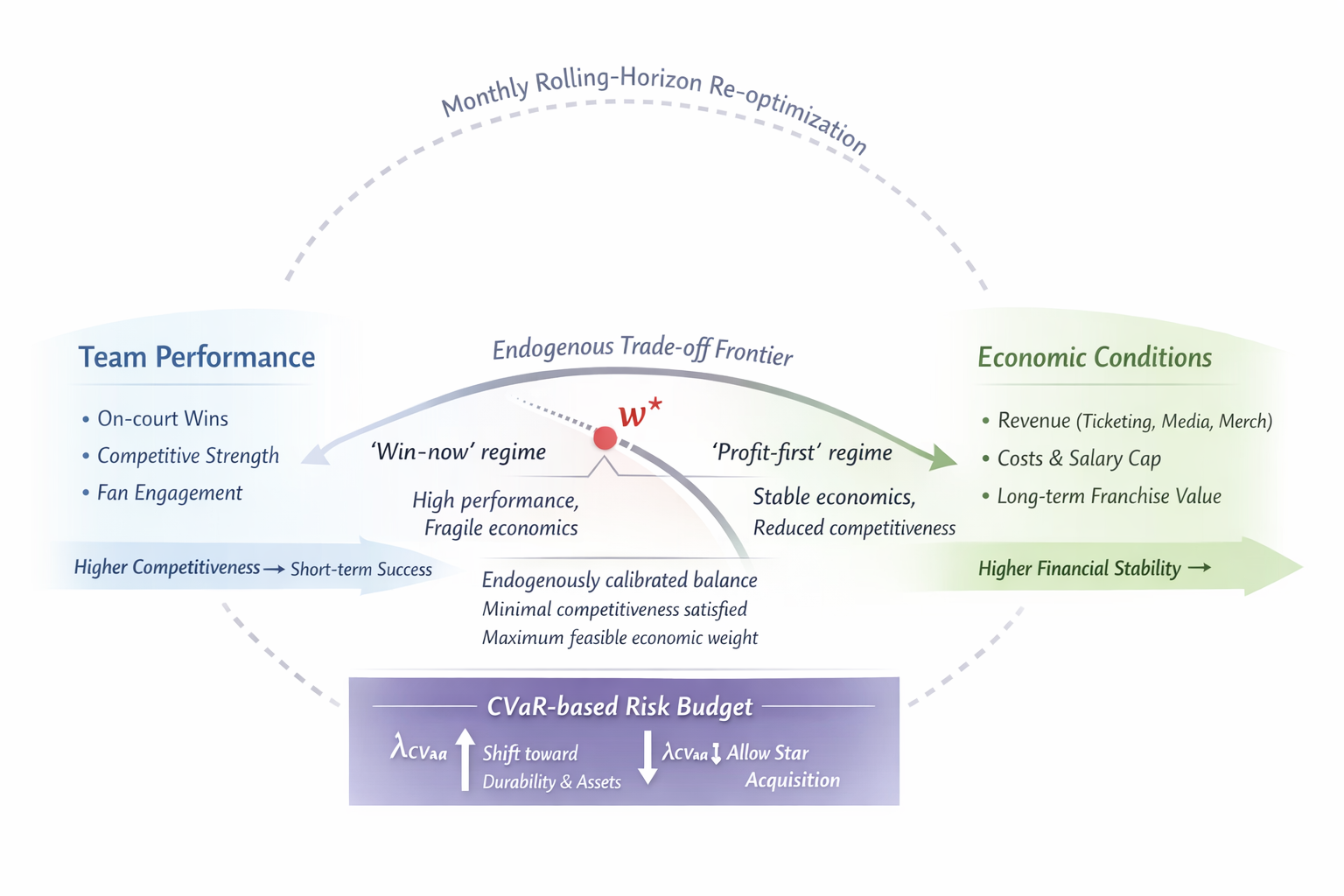}
    \caption{The relationship between team performance and economic conditions}
    \label{fig:The relationship between team performance and economic conditions}
\end{figure}

\hspace{1.4em} At each decision epoch $t$, the engine selects a vector of decisions $x_t$ spanning six modules: (1) ticket pricing ($x_{tick,t}$: e.g. base price, season discount, premium seat markup), (2) marketing and sponsorship ($x_{merc,t}$: e.g. merchandise pricing, number of sponsors, streaming rights adjustments), (3) venue operations ($x_{ven,t}$: e.g. advertising slots, maintenance budget, concessions pricing), (4) player acquisitions ($x_{acq,t}$: e.g. pursue a marquee star $y_{star,t}\in\{0,1\}$; set rookie development rate $r_{rookie,t}$), (5) trades ($x_{trade,t}$: e.g. number of stars traded $n_{st,t}$, draft picks kept $n_{pick,keep,t}$ or acquired $n_{pick,up,t}$), and (6) player contracts ($x_{cont,t}$: e.g. offer extensions $y_{ext,t}$, minimum deals $n_{min,t}$, bonus incentives $r_{bonus,t}$). Together, these constitute the control vector $x = \{x_t\}_{t=0}^{H-1}$. 

 The \textbf{distributionally robust objective} is to maximize the worst-case expected total discounted franchise value, combining financial profits and on-court performance (plus terminal value). Formally, we solve: 
\[ 
\max_{x} \;\min_{P\in\mathcal{P}} \; E_{P}\Bigg[ \sum_{t=0}^{H-1} \gamma^t \Big( w\, Profit_t(x) \;+\; (1-w)\,Perf_t(x) \Big)\;+\; Val_{long,H}(x) \Bigg], 
\] 
where $0 < w < 1$ balances financial vs. competitive objectives. Rather than fixing $w$ arbitrarily, we calibrate it to the smallest value $w^*$ that ensures a minimum competitiveness level (e.g. $Perf_t$ stays above $S_{\min}=0.48$ over any 24-month window). This approach guarantees the franchise's competitive mandate; $w^*$ is found via a one-dimensional search outside the main MIP and then treated as a parameter in the optimization. 

 $\mathcal{P}$ denotes the ambiguity set of probability distributions for uncertainties. We construct $\mathcal{P}$ in a moment-based DRO fashion: for each month $t$, let $Z_t$ be the vector of uncertain factors (e.g. macro multipliers, injury indicators). We define 
\[ \mathcal{P}_t = \Big\{ P:\; |E_P[Z_t] - \hat{\mu}_t| \le \rho_{\mu,t},\;\; \| \mathrm{Var}_P(Z_t) - \hat{\sigma}_t^2 \| \le \rho_{\sigma,t} \Big\}, \] 
where $\hat{\mu}_t, \hat{\sigma}_t$ are nominal mean and standard deviation (from data or scenarios), and $\rho_{\mu,t}, \rho_{\sigma,t}$ are allowable deviation bounds. The overall ambiguity set is $\mathcal{P} = \prod_{t=0}^{H-1} \mathcal{P}_t$. Thus, the inner $\min_{P\in\mathcal{P}} E_P[\cdot]$ represents a worst-case expectation over all distributions consistent with those moment constraints, protecting the solution against adverse distributional shifts.

\noindent \textbf{Constraints:} For feasibility and policy compliance, we impose the following constraints at each month $t$: 
\[ 
\begin{cases}
Cash_{t-1} + Profit_t \ge C_{min}, &\text{(liquidity buffer)}\\ 
Debt_t \le D_{max}, &\text{(solvency limit)}\\ 
\displaystyle \frac{EBITDA_t}{DebtService_t} \ge \psi_{DSCR}, &\text{(debt service coverage)}\\ 
\text{TradeRules}(x_t;\theta_{rules}) = \text{True}, &\text{(league trade regulations)}\\ 
13 \;\le\; |I_t| \;\le\; 18, &\text{(roster size limits)}
\end{cases}
\] 
These ensure the team maintains a cash reserve ($C_{min}$), keeps debt under $D_{max}$, satisfies a debt-service coverage ratio $\psi_{DSCR}$, obeys NBA trade rules (via the predicate $\text{TradeRules}$ with parameters $\theta_{rules}$), and maintains a legal roster size (13--18, effectively 14--15 active players). \textbf{Tail-Risk (CVaR) Constraint:} To curb injury-related downside risk (large ``dead money'' salary losses), we impose a CVaR budget. Let $\Omega$ be the scenario set and $\alpha=0.95$. Introduce $\zeta_t$ as the $\alpha$-quantile (VaR) of injury-related salary loss in month $t$, and $\nu^\omega_t$ as the loss exceeding $\zeta_t$ in scenario $\omega$. We require for each $t$: 
\[ 
\zeta_t + \frac{1}{(1-\alpha)|\Omega|}\sum_{\omega\in\Omega} \nu^\omega_t \;\le\; \eta\, Cap_t, 
\] 
\[ 
\nu^\omega_t \;\ge\; \sum_{i\in I_t} Sal_{i,t}\, I\!\big(Dur_{i,t}^\omega < \tau\big)\;-\;\zeta_t,\qquad 
\nu^\omega_t \ge 0, \;\forall \omega\in\Omega~, 
\] 
ensuring the Conditional Value-at-Risk (CVaR$_\alpha$) of lost salaries does not exceed $\eta$ times the cap (we use $\eta=0.25$). Here $I(Dur_{i,t}^\omega < \tau)$ is an indicator that player $i$'s durability in scenario $\omega$ falls below threshold $\tau$ (meaning $i$ is out injured and $Sal_{i,t}$ becomes ``dead money''). This linear formulation introduces slack variables $\nu^\omega_t$ for tail losses and keeps their average excess (scaled by $1-\alpha$) within $\eta\,Cap_t$. By tuning $\eta$, management sets the acceptable worst-case injury cost (25\% of cap here). All such parameters (e.g. $C_{min}, D_{max}, \psi_{DSCR}, \theta_{rules}, \tau$) are inputs.

\subsection{Solution Algorithm: Proximal Bundle Method}

\hspace{1.4em} Solving the full robust RH-SMIP is challenging due to its scale, integer roster decisions, and non-smooth DRO/CVaR terms. We apply a combination of Lagrangian relaxation and a proximal bundle method, embedded in a rolling-horizon simulation. The procedure is outlined below:

\begin{algorithm}[H]
\caption{NYK-ADMS Decision Engine}
\label{alg:Bundle}
\begin{multicols}{2}
\begin{algorithmic}[1]

\State \textbf{Input:}
\WrapState{Current state $S_t$; planning horizon $H$; scenario set $\Omega$; risk level $\alpha=0.95$; CVaR budget $\eta=0.25$; performance floor parameters $T_0=24$, $S^{\min}=0.48$.}

\State \textbf{Scenario generation:}
\WrapState{Generate ambiguity-aware scenario set $\Omega$.}

\State \textbf{Calibrate weight $w$:}
\WrapState{Grid search $w\in[0,1]$ to enforce performance floor; take smallest feasible $w$.}

\State \textbf{Initialization:}
\WrapState{Dual multipliers $\bm{\lambda},\bm{\mu}$; empty bundle $\mathcal{B}\leftarrow\emptyset$; proximal weight $u$; tolerance $\epsilon$.}

\For{$k=1$ to $K_{\max}$}
  \State Solve decomposed subproblems to get $\bm{x}^{(k)}$
  \State Compute subgradient $g_k$ (from constraint violations and linearization error $e_k$)
  \State Update bundle: $\mathcal{B}\leftarrow \mathcal{B}\cup\{(g_k,e_k)\}$
  \State Solve bundle master QP to get search step $\bm{d}_k$
  \State Update duals: $\bm{\lambda}^{(k+1)}\leftarrow \bm{\lambda}^{(k)} + \bm{d}_k$
  \If{$\|\bm{d}_k\|<\epsilon$}
    \State \textbf{Terminate:} convergence achieved
  \EndIf
\EndFor

\State \textbf{Feasibility pump:}
\WrapState{Repair integrality and enforce all core constraints.}

\State \textbf{Output:}
\WrapState{Monthly decision $\bm{x}^{*}_t$ and diagnostics (e.g. CVaR dual $\lambda_{\text{CVaR}}$).}

\State \textbf{Roll horizon:}
\WrapState{Update state to $S_{t+1}$ and repeat for next month.}

\end{algorithmic}
\end{multicols}
\end{algorithm}

\subsection{Statistical Analysis of Optimization Outcomes}
\begin{figure}[H]
    \centering
    \includegraphics[width=0.9\linewidth]{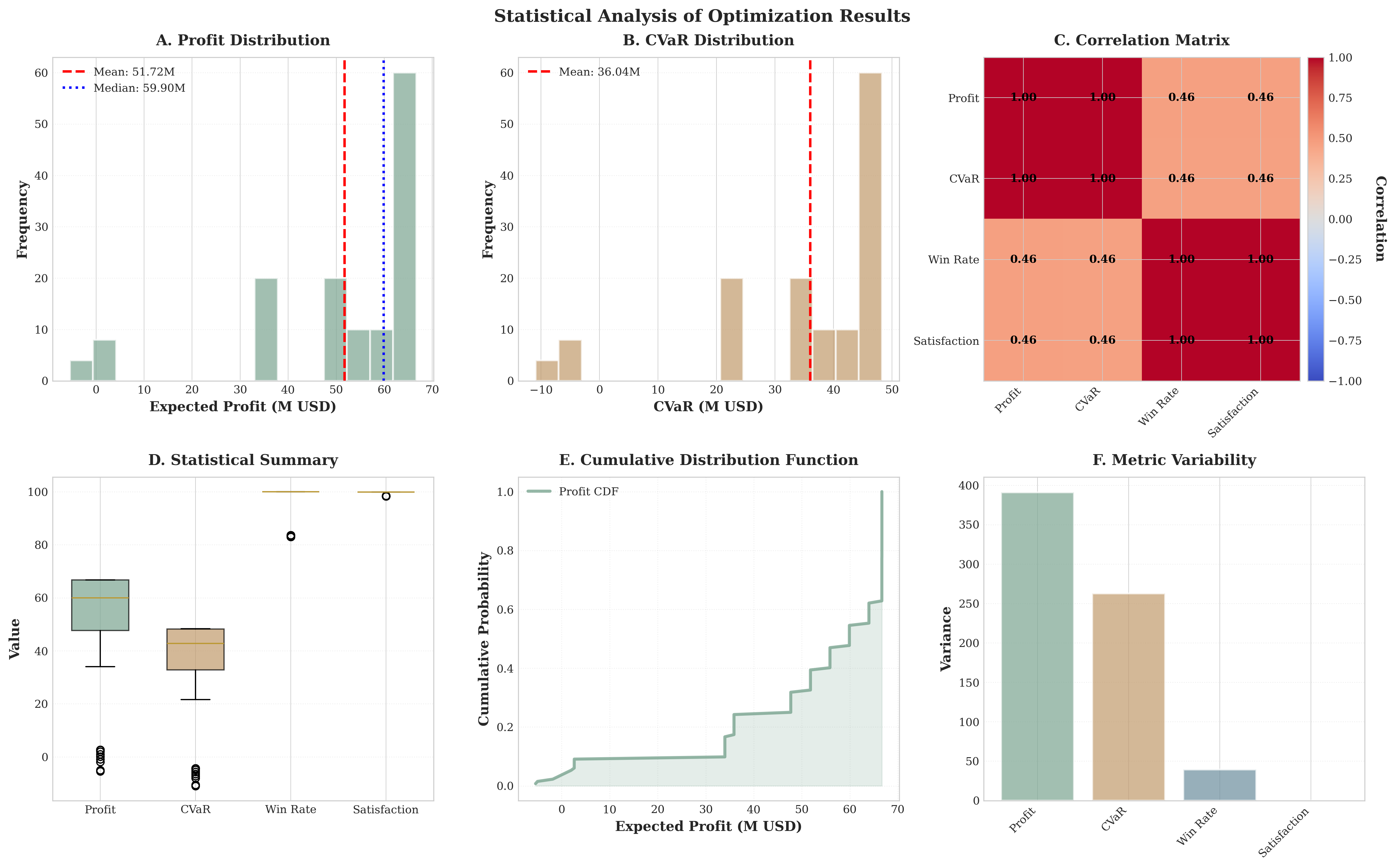}
    \caption{Statistical Analysis of Optimization Outcomes}
    \label{fig:Statistical Analysis of Optimization Outcomes}
\end{figure}

\hspace{1.4em} Panels (A)--(F) summarize the statistical properties of model-generated solutions across all feasible configurations. Panels (A) and (B) report the empirical distributions of expected profit and CVaR, respectively, while Panel (E) presents the corresponding cumulative distribution. Panel (C) shows the correlation structure among financial and competitive metrics. Panel (D) provides a boxplot-based summary of dispersion and outliers, and Panel (F) compares cross-metric variability. Together, these panels characterize the distributional shape, tail behavior, coupling structure, and robustness of the optimization outcomes.

The optimization results exhibit a multi-regime profit distribution, with the median exceeding the mean, indicating left-tail risk driven by a small number of loss-tolerant solutions. CVaR displays a similar structure, suggesting that most feasible strategies maintain strong downside protection despite isolated adverse outcomes. The correlation matrix reveals a strong alignment between profit and CVaR, while their association with competitive performance remains moderate, implying partial decoupling between financial optimization and on-court success. Boxplots and variance comparisons further show that uncertainty is concentrated in financial metrics, whereas win rate and satisfaction remain tightly bounded. Overall, the results indicate that the model achieves financially robust solutions without compromising competitive stability, consistent with its risk-constrained design.

\subsection{Interim Interpretation}
\begin{figure}[htbp]
    \centering
    \includegraphics[width=0.6\linewidth]{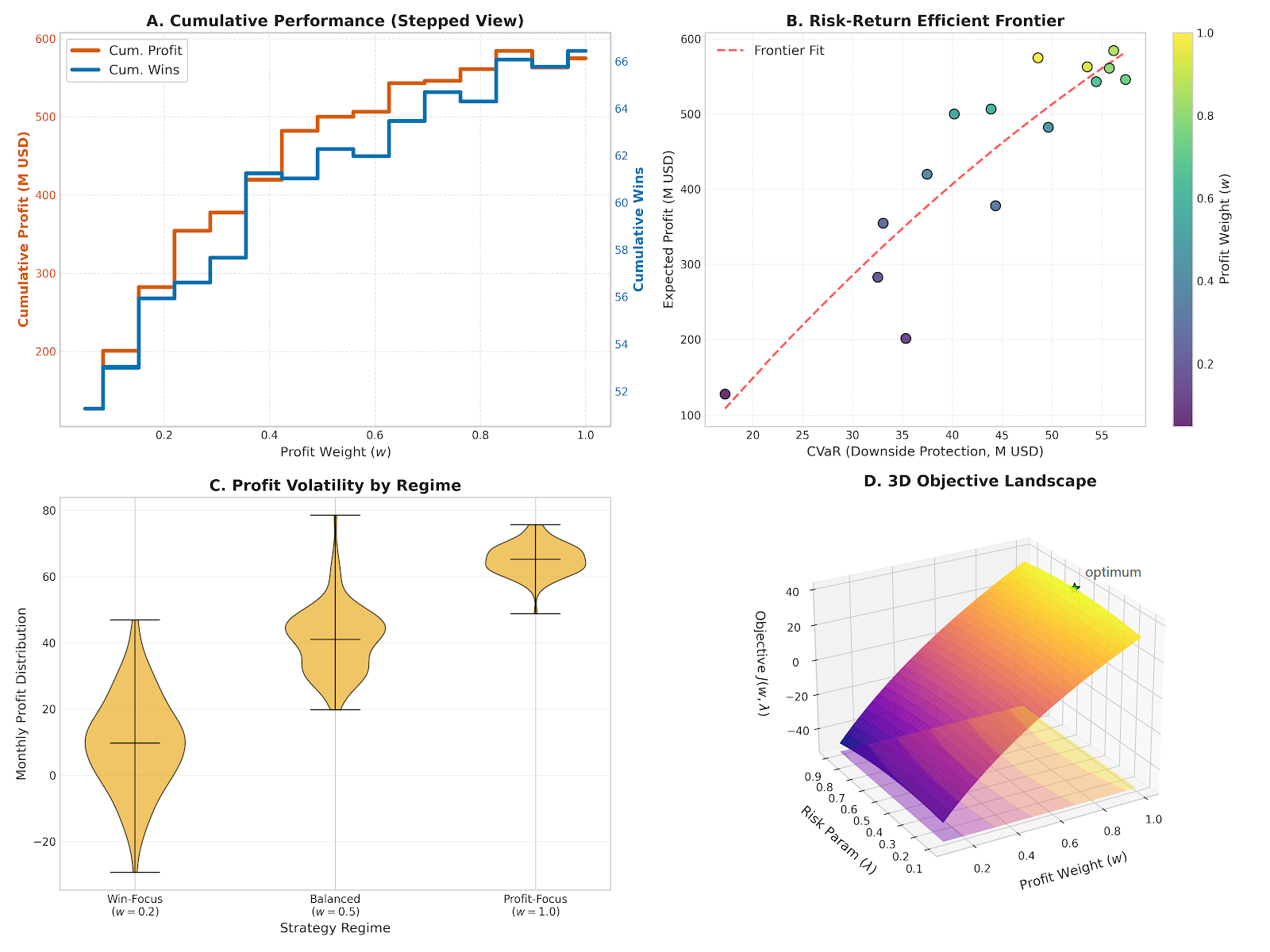}
    \caption{The performance of the model}
    \label{fig:the performance of the model}
\end{figure}

\hspace{1.4em}The performance results above demonstrate a clear synergy between financial and competitive outcomes: as the profit weight $w$ increases, both cumulative profit and cumulative wins rise in a near-monotonic and approximately linear manner. This indicates that strengthening commercial emphasis does not crowd out on-court success, but instead enhances overall performance through improved financial capacity and operational support. The risk--return profile and the profit--risk efficient frontier jointly reveal a well-defined Pareto structure, with a distinct regime shift from loss-tolerant solutions at $w=0$ to self-financing solutions for $w \ge 0.1$, where both expected profit and CVaR become positive. Moreover, the composite objective $J(w)$ is strictly increasing in $w$, and the full log consistently selects $w^\ast = 1.0$ across all rolling-horizon months, establishing commercial optimization as the binding managerial priority.

Under the profit-dominant regime implied by $w^\ast = 1.0$, the model prescribes a revenue-driven operating strategy: \textbf{(1)} Ticketing should adopt moderate base discounts ($p_{\text{reg}} = 0.50$) to expand attendance while simultaneously maintaining a high-end premium tier ($m_{\text{prem}} = 3.0$) to capture surplus from fans with high willingness to pay, avoiding additional season-ticket discounts that could erode long-term revenue. \textbf{(2)} Commercial income is maximized through full monetization of merchandise ($p_{\text{merc}} = 1.0$), aggressive sponsorship acquisition ($n_{\text{spon}} = 5$), and active deployment of streaming channels ($r_{\text{stream}} = 1.0$). \textbf{(3)} Venue operations emphasize growth-oriented advertising intensity ($n_{\text{ads}} = 3$) combined with controlled maintenance expenditure ($b_{\text{maint}} = 0.50$), generating sustained positive cash flow while team operations prioritize cost-efficient roster depth rather than high-risk star acquisitions.
\section{Transaction Layer: Acquisition, Free Agency, and Trade}

\subsection{The Dual-Engine Handshake Protocol}
\begin{figure}[htbp]
    \centering
    \includegraphics[width=0.8\linewidth]{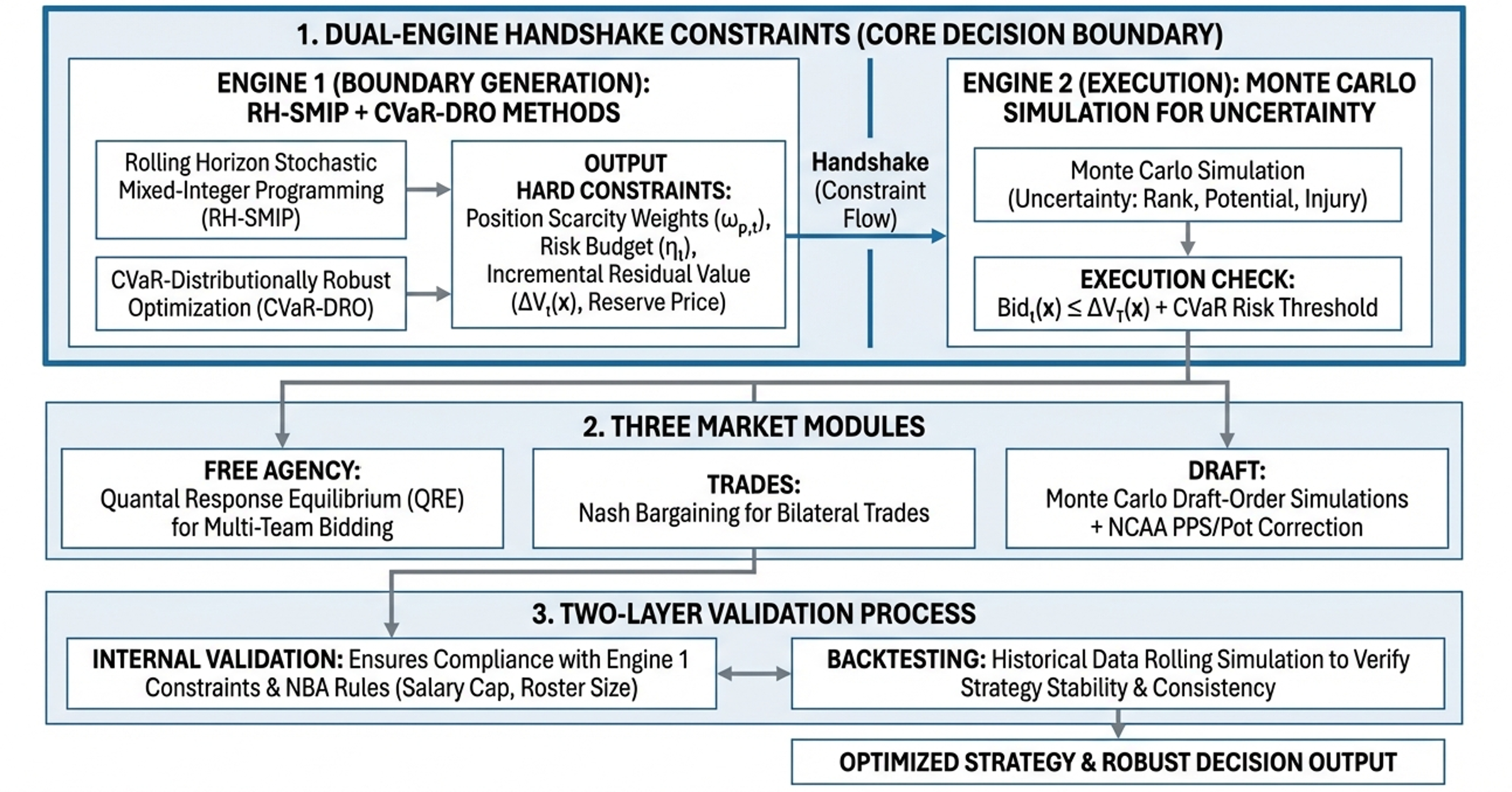}
    \caption{Dual-engine handshake constraint system: Generating rational boundaries for market execution.}
    \label{fig:Dual-engine handshake constraint system}
\end{figure}

The transaction layer operationalizes the long-horizon valuation engine by introducing a \textbf{market execution layer}. Instead of treating acquisition as an unconstrained optimization problem, we construct a hierarchical ``handshake'' in which Engine 1 (valuation) generates hard rationality boundaries for Engine 2 (execution).
Let $\mathcal{R}_t$ be the roster state and $\eta_t$ be the risk budget inherited from the core optimization layer. We define the \textbf{Incremental Surplus} $\Delta V_t(x)$ of any asset $x$ and the \textbf{Admissible Acquisition Set} $\Phi_t$ as:
\begin{equation}
\label{eq:handshake_set}
\Phi_t = \left\{ x \in \mathcal{F}_t \ \Bigg| \ 
\begin{aligned}
&\text{Bid}_t(x) \le \Delta V_t(x) \equiv V_t(\mathcal{R}_t \oplus x) - V_t(\mathcal{R}_t) \quad \text{(Rationality Anchor)} \\
&\text{CVaR}_{\alpha}\left[-\Delta \widetilde{V}_t(x)\right] \le \eta_t \quad \text{(Risk Veto)}
\end{aligned}
\right\}
\end{equation}
This formulation ensures defense against the "Winner's Curse" by strictly rejecting transactions with negative expected surplus or excessive tail risk, regardless of market pressure.

\subsection{Market Modules: Draft, Agency, and Trade}

We deploy three specialized operators to search for the optimal action $x^* = \arg\max_{x \in \Phi_t} \mathbb{E}[\Delta V_t(x)]$ within the admissible set.

\paragraph{(I) Draft Module: Bayesian Prospect Valuation} 
We adapt the core player-potential model for NCAA prospects by introducing a rank-based prior. Let $r_i$ be the scout rank. The adjusted potential $Pot^{NCAA}_i$ combines performance (PPS) and upside priors:
\begin{equation}
Pot^{NCAA}_i = 100 \cdot I_i \cdot \left[ (1-\lambda_i)PPS^{NCAA}_i + \lambda_i \cdot \text{clip}\left( c_1(r_i+1)^{-\alpha} - c_2 PPS^{NCAA}_i \right) \right]
\end{equation}
where $\lambda_i$ weights the age-dependent upside, and $(r_i+1)^{-\alpha}$ models the power-law decay of draft value.

\paragraph{(II) Free Agency: QRE and the "New York Premium"} 
We model multi-team bidding via a \textbf{Quantal Response Equilibrium (QRE)}. The probability of player $i$ signing with team $j$ is $P_{ij} = e^{\tau U_{ij}} / \sum_k e^{\tau U_{ik}}$. Solving for the indifference condition $U_{i,NYK} = U_{i,small}$ reveals the structural \textbf{Market Size Premium}:
\begin{equation}
\label{eq:ny_premium}
S^*_{i,NYK} = S_{i,small} \cdot \exp\left( -\frac{\beta_{mkt}}{\beta_{sal}} (\text{Mkt}_{NYK} - \text{Mkt}_{small}) \right) < S_{i,small}
\end{equation}
This analytic result quantifies that the Knicks can offer lower salaries for identical utility, generating a "cap surplus" of approx. \$4--6M/year, which is reinvested into roster depth $\mathcal{R}_{depth}$ to minimize risk.

\paragraph{(III) Trade Module: Nash Bargaining} 
Bilateral trades are resolved via the asymmetric Nash Bargaining Solution, maximizing the joint surplus product subject to Pareto admissibility:
\begin{equation}
T^* = \arg\max_{T \in \Phi_t} \left( \Delta V^{NYK}_t(T) \right)^{\theta} \left( \Delta V^{Opp}_t(T) \right)^{1-\theta}, \quad \text{s.t. } \Delta V^{Opp}_t(T) \ge 0
\end{equation}
This ensures that any accepted trade strictly increases the robust objective value for the Knicks while satisfying the counterparty's participation constraint.

\subsection{Risk-Adjusted Market Outcomes}

The model gives three trade plans:
\begin{enumerate}
    \item[\textbf{Trade~1}] \textbf{Counterparty}: Oklahoma City Thunder \newline
          \textbf{Traded Away}: Dillon Jones; Trey Jemison III \newline
          \textbf{Acquired}: Luke Kennard (\$4.0M) \newline
          \textbf{Draft Pick}: (Keep) 2026 1st (\#21)(select Qayden Samuels SF(expected\#22)); (Receive) 2027 2nd \newline
          \textbf{Cash}: \$1.0M \newline
          $\Delta$objective: +0.037

    \item[\textbf{Trade~2}] \textbf{Counterparty}: Utah Jazz \newline
          \textbf{Traded Away}: Jordan Clarkson; Kevin McCullar Jr. \newline
          \textbf{Acquired}: Royce O'Neale (\$9.0M) \newline
          \textbf{Draft Pick}: (Keep) 2026 1st (\#21)(select Qayden Samuels SF(expected\#22)); (Receive) 2028 2nd \newline
          \textbf{Cash}: \$0.5M \newline
          $\Delta$objective: +0.007

    \item[\textbf{Trade~3}] \textbf{Counterparty}: San Antonio Spurs \newline
          \textbf{Traded Away}: Mitchell Robinson; Ariel Hukporti \newline
          \textbf{Acquired}: Stretch 5 (\$12.0M) \newline
          \textbf{Draft Pick}: (Keep) 2026 1st (\#21)(select Qayden Samuels SF(expected\#22)); (Receive) 2027 2nd \newline
          \textbf{Cash}: \$0.5M \newline
          $\Delta$objective: +0.016
\end{enumerate}

\begin{figure}[H]
    \centering
    \includegraphics[width=0.5\linewidth]{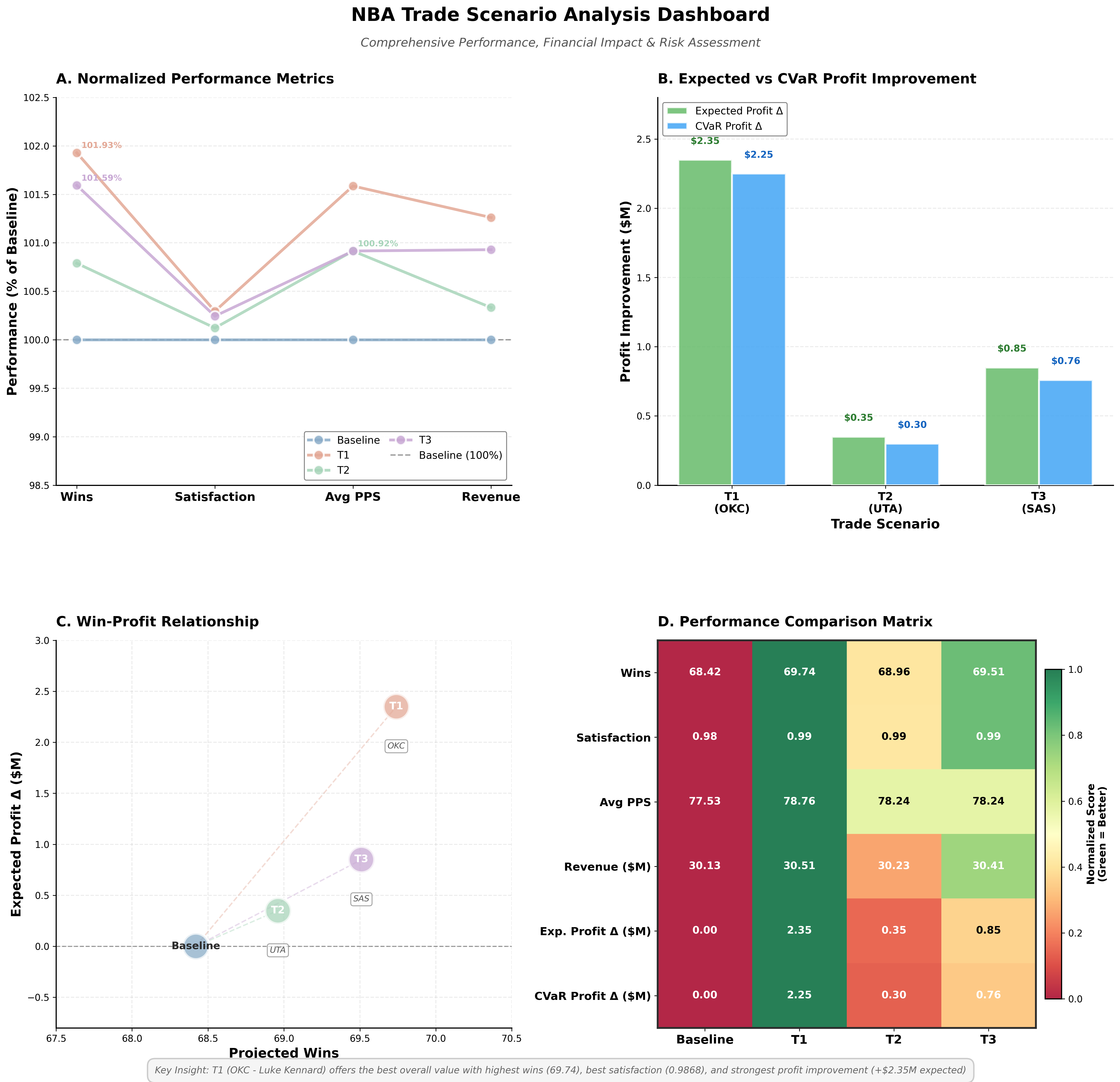}
    \caption{The benefits of the trades}
    \label{fig:The benefits of the trade}
\end{figure}
Based on the comparative outcomes between the baseline and the three acquisition
scenarios (T1--T3), the transaction layer can be interpreted as the executable face
of the core valuation model. All scenarios preserve roster size, competitive
performance, and win saturation (wins $\approx 69$, $S_{\text{sat}}\approx 0.986$), ensuring
that on-court stability is not sacrificed for short-term arbitrage. Among them,
T1 delivers the largest improvement in expected monthly profit while simultaneously
improving CVaR, yielding the highest increase in the composite objective. In contrast,
T2 and T3 generate only marginal gains once risk adjustments are applied. The
important point is not that the model finds ``more trades,'' but that it restricts
execution to actions consistent with reservation values and the franchise-wide CVaR
budget, thereby accepting only value-accretive, risk-feasible transactions and
rationally rejecting overpriced or tail-risk-dominated acquisitions in a competitive market.

Unfortunately, based on our investigation, Dillon Jones is deeply beloved by both Knicks players and fans. Any trade involving him would inevitably impact player morale and team culture.
\section{External Structural Shock I: League Expansion}
\begin{figure}[htbp]
    \centering
    \includegraphics[width=0.6\linewidth]{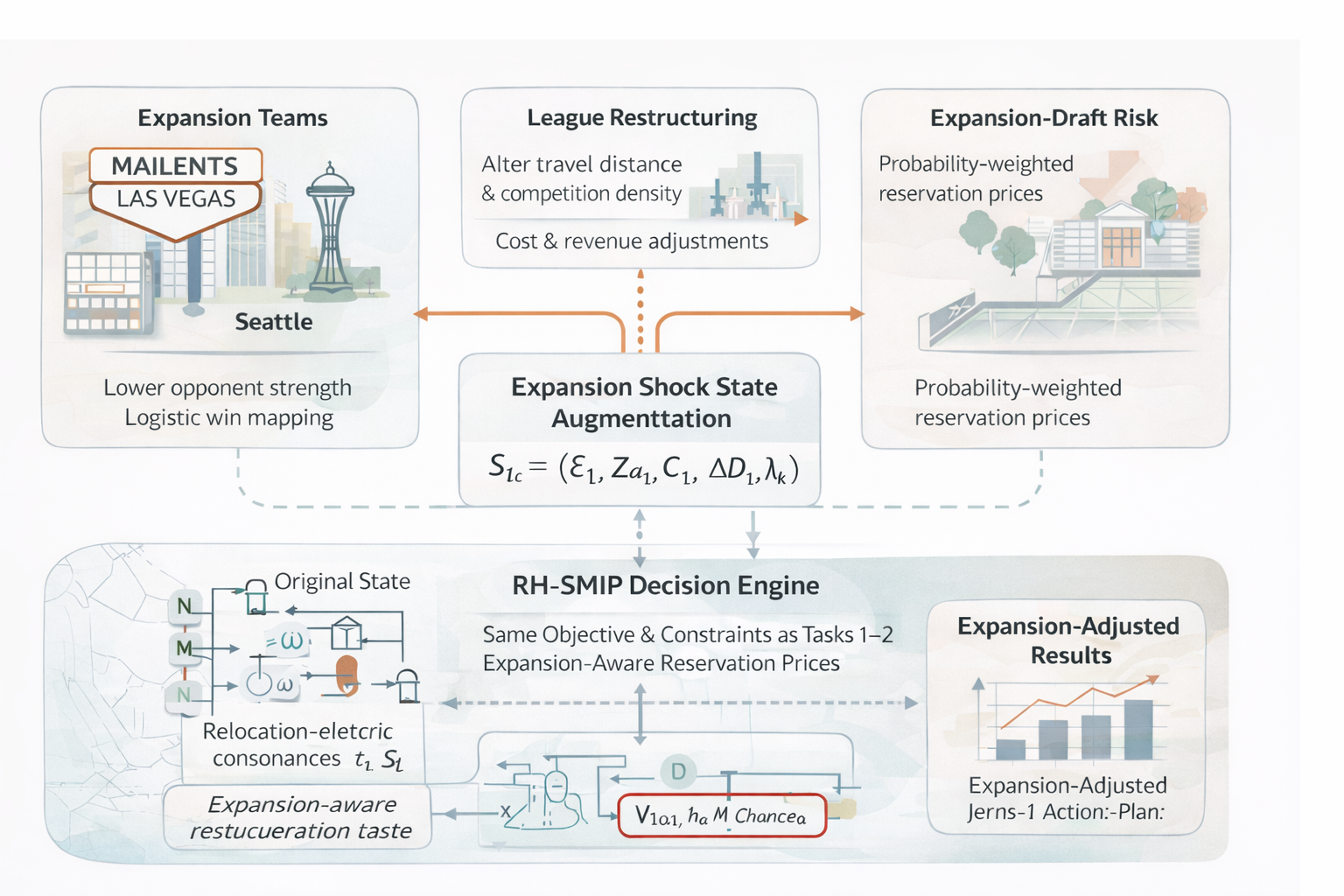}
    \caption{Shock-propagation structure for league expansion}
    \label{fig:The model of task~3}
\end{figure}
\subsection{Expansion Shock Modeling}

We model the league expansion not as a new optimization problem, but as a \textbf{Structural Perturbation Operator} $\mathcal{T}_{\text{exp}}$ applied to the RH-SMIP state space. The expansion-augmented state dynamics are governed by a unified shock system $\mathcal{S}_t(\mathcal{C}) = \langle \Delta D_t, \Phi^{comp}_t, \delta^{weak}_y \rangle$, which simultaneously transforms competitive topology, financial flows, and asset valuation.

The operational adjustments are rigorously formalized as a coupled system of equations:

\begin{equation}
\label{eq:expansion_shock}
\begin{aligned}
\textbf{I. Competitive Damping:}\quad &
p_{\text{win}}(t) = \sigma \left( \beta_1 \left[ TS^{\text{NYK}}_t - \underbrace{(1 - \mathbb{I}_{\text{exp}}\delta_y)\overline{TS}^{\text{opp}}_t}_{\text{Diluted Strength}} \right] + \beta_2 H \right) \\
\textbf{II. Cost-Revenue Affine:}\quad &
\left[ \begin{matrix} \mathbf{Cost}_t^{\text{exp}} \\ \mathbf{Rev}_t^{\text{exp}} \end{matrix} \right]
=
\left[ \begin{matrix} \mathbf{Cost}^0_t \\ \mathbf{Rev}^0_t \end{matrix} \right]
+
\underbrace{\left[ \begin{matrix} c_D & 0 \\ 0 & \xi_{nat} \end{matrix} \right] \left[ \begin{matrix} \Delta D_t \\ \Delta N/N_0 \end{matrix} \right]}_{\text{Structural Drift}}
+
\underbrace{\mathcal{M}(\Phi^{comp}_t)}_{\text{Density Elasticity}} \\
\textbf{III. Risk-Adjusted Valuation:}\quad &
\Delta V^{\text{exp}}_t(x) = \sum_{i \in \mathcal{R} \oplus x} \underbrace{(1 - p_{\text{pick}}\mathbb{I}_{i \notin \text{Prot}})}_{\text{Retention Prob.}} V_i - V^{\text{exp}}_t(\mathcal{R})
\end{aligned}
\end{equation}

where $\sigma(\cdot)$ is the logistic sigmoid, $\delta_y \in \{0.3, 0.2\}$ represents the exogenous weakness of expansion teams in their first two seasons, and $\mathcal{M}(\cdot)$ encapsulates the nonlinear elasticity of staff costs and local revenues to competition density $\Phi^{comp}$. This formulation ensures that the expansion draft risk (Eq. III) acts as a hard "handshake constraint" on player acquisition, rejecting high-value assets that cannot be protected.

\subsection{Unified Solution and Spatial-Structural Adjustment}

We unify the league expansion shock into a spatial-structural operator $\mathcal{T}_{\mathcal{C}}$ acting on the RH-SMIP engine. The scenario set $\mathcal{C} \in \{\text{BOS}, \text{NYC}, \text{MIN}\}$ is explicitly defined as: (1) \textbf{BOS}: Relocation of the Minnesota Timberwolves to Boston (creating a dual-team market analogous to Los Angeles); (2) \textbf{NYC}: Establishment of a new expansion franchise directly within New York City (intra-city co-location); and (3) \textbf{MIN}: Administrative reassignment of the Timberwolves to the Eastern Conference.

To rigorously quantify the propagation of these topological shocks, we formalize the \textbf{Location Impact Path} as follows:

\[
\resizebox{0.97\linewidth}{!}{$
\mathcal{C}_t
\xrightarrow{\text{Geo-Structure}}
\left[ \begin{matrix} \Delta D_t \\ \overline{TS}^{opp}_t(\mathcal{C}_t) \\ N^{near}(\mathcal{C}_t) \end{matrix} \right]
\xrightarrow{\text{Mechanism}}
\left[ \begin{matrix} \Phi^{comp}(\mathcal{C}_t) \\ Cost^{travel}_t \\ p_{win}(t) \end{matrix} \right]
\xrightarrow{\text{Fin-Comp}}
\left[ \begin{matrix} \text{Module Cost/Rev Formulas} \\ \text{Competitive Outcomes} \end{matrix} \right]
\xrightarrow{\text{Optimizer}}
\left[ \begin{matrix} V^{exp}(\mathcal{R}) \\ \Delta x^* \end{matrix} \right]
$}
\]

We parameterize the location impact via a \textbf{Spatial Topology Vector} $\mathbf{z}_{\mathcal{C}} = [\Delta D(\mathcal{C}),\, \delta_{\Phi}(\mathcal{C})]^T$, where $\Delta D$ is the travel radius perturbation and $\delta_{\Phi} = (N^{near}-N^{near}_0)/N^{near}_0$ represents the local competition density shift. Defining the \textbf{Revenue Elasticity Vector} $\boldsymbol{\psi} = [\psi_{rsn}, \psi_{str}, \psi_{gate}, \psi_{s}] = [0.35, 0.20, 0.15, 0.25]$ and global dilation coefficient $\xi_{nat}=0.06$, the expansion-adjusted Profit Function $\Pi_t^{\mathcal{C}}$ is derived via the following compact topology mapping:

\begin{equation}
\label{eq:spatial_map}
\begin{aligned}
\textbf{Structure:}\quad & \Phi^{comp}(\mathcal{C}) = 1 + \lambda_{comp} \cdot \delta_{\Phi}(\mathcal{C}), \quad \text{with } \lambda_{comp}=0.20 \\
\textbf{Cost:}\quad & Cost^{ops}_t(\mathcal{C}) = Cost^{ops,0}_t \big[1 + 0.05(\Phi^{comp}(\mathcal{C})-1)\big] + \underbrace{c_D \Delta D(\mathcal{C})}_{\text{Travel Penalty}} \\
\textbf{Revenue:}\quad & \mathbf{Rev}_t(\mathcal{C}) = \mathbf{Rev}^0_t \odot \Big( \mathbf{1} + \boldsymbol{\xi} \frac{\Delta N}{N_0} - \boldsymbol{\psi} \odot \mathbf{I}_{loc} (\Phi^{comp}(\mathcal{C}) - 1) \Big)
\end{aligned}
\end{equation}

where $\odot$ denotes the Hadamard product, and $\mathbf{I}_{loc}$ isolates density-sensitive streams. This formulation explicitly quantifies how the \textbf{NYC} scenario maximizes $\Phi^{comp}$ (triggering revenue cannibalization $\boldsymbol{\psi}$), whereas \textbf{BOS} and \textbf{MIN} shifts primarily impact the travel penalty $c_D$ and competitive win probabilities $p_{win}(t)$.

The unified solver produces the expansion-adapted policy $x^{*,exp}$ by optimizing the structurally perturbed objective. The strategic recourse is decomposed as:
\[
\Delta x^{*}_{26/27}(\mathcal{C}) = \arg\max_{x \in \mathcal{X}(\mathcal{C})} \mathbb{E}_{\mathbb{P}}\left[ \sum_{t} \gamma^t \Pi_t^{\mathcal{C}}(x) \right] - x^{*,base},
\]
identifying the precise marginal adjustments in ticketing $\Delta x_{tick}$ and acquisition aggressiveness $\Delta x_{acq}$ required to immunize the franchise against specific location-induced dilution.

\subsection{Problem Solving: Strategy Adjustment under League Expansion}

\hspace{1.4em}
Using the RH-SMIP decision engine and transaction layer introduced above, we evaluate how
league-determined structural changes induced by expansion propagate through the
Knicks' optimal in-season strategy.
All scenarios are solved under an identical rolling-horizon configuration with
unchanged salary cap, roster size, scheduling rules, and risk parameters.
Across all scenarios, the calibrated objective weight remains unchanged
($w^*=0$ in the md outputs), indicating that league expansion does not alter the
underlying managerial regime.
Instead, its effects materialize through binding system constraints and league-wide
revenue reallocation, rather than through a shift in managerial preferences.

Quantitatively, expansion generates a persistent negative profit shock whose magnitude
depends critically on \emph{where} the new franchise is placed.
Relative to the baseline (\$200.6M), all three expansion configurations reduce cumulative
profit: BOS relocation is the least harmful ($-84.5$M, $-42.1\%$), NYC co-location is the
most harmful ($-104.3$M, $-52.0\%$), and MIN reassignment falls in between
($-94.4$M, $-47.1\%$).
The ranking aligns closely with the competition-density multiplier
$\phi_{\text{comp}}$ and travel increments, indicating that local market overlap
(NYC co-location) dominates purely geographic frictions.
This confirms that expansion shocks are primarily driven by revenue cannibalization
rather than by on-court competition effects alone.

From a managerial perspective, the key insight is not the magnitude of the shock itself,
but its \emph{non-arbitrageable} nature.
The optimal policy exhibits near-zero adjustments in ticket pricing, media investment,
sponsorship intensity, and roster transactions across all scenarios.
This invariance is not a failure of responsiveness, but a diagnostic outcome of the model:
within feasibility and risk boundaries, a single franchise lacks effective control levers
to offset league-level structural dilution.
Accordingly, the optimal response is strategic stability rather than reactive intervention.

Expansions proximate to large media markets (NYC co-location) impose disproportionately
large financial externalities through demand cannibalization, while distant relocations
(BOS $\rightarrow$ Seattle) generate substantially smaller impacts that are largely absorbed
by existing operating margins.
The RH-SMIP framework thus serves not to recommend aggressive countermeasures, but to
certify the limits of managerial adjustment and to prevent overreaction to shocks that
cannot be mitigated at the team level.

\section{External Structural Shock II: Media Rights Transition}

We model media strategy not as an isolated marketing lever, but as a \textbf{Regime-Switching Control Problem} embedded within the RH-SMIP. The objective is to optimize the \textbf{Streaming Intensity Vector} $x_{med,t} = [r_{stream,t}, n_{spon,t}]^\top$ to navigate the transition from legacy Regional Sports Network (RSN) contracts to Direct-to-Consumer (DTC) platforms.

\subsection{Coupled Revenue-Risk Topology}
We construct a unified state-space model coupling the media portfolio profit $\Pi^{med}_t$ with subscriber dynamics and cross-channel cannibalization. The optimizer solves for $r_{stream,t}$ subject to a \textbf{Risk-Gated Hamiltonian}:

\vspace{-1.2em} 
\begin{subequations}
\small 
\begin{align}
    \label{eq:media_profit}
    \max_{\{r_{stream,t}\}} \;\; \Pi^{med}_t &= 
    \underbrace{F_{RSN} e^{-\lambda_{cut} t}}_{\text{\textbf{Legacy Decay}: Fixed Contract}} 
    \;+\; 
    \underbrace{\mu(r_{stream,t}) \cdot S_{sat}(W_t) - C_{tech}}_{\text{\textbf{DTC Growth}: Intensity-Driven}} 
    \;-\; 
    \underbrace{\xi_{can} \cdot r_{stream,t} \cdot Gate_t}_{\text{\textbf{Cannibalization Penalty}}} \\
    \label{eq:risk_gate}
    \text{s.t.} \;\; \Delta r_{stream,t} &> 0 \iff 
    \underbrace{\frac{\partial \mathbb{E}[\Pi^{med}_t]}{\partial r_{stream,t}} > 0}_{\text{Positive Marginality}} 
    \;\land\; 
    \underbrace{\mathrm{CVaR}_\alpha(-\Pi_t) \le \eta \mathrm{Cap}_t}_{\text{Risk Admissibility}}
\end{align}
\end{subequations}
\vspace{-0.5em}

where $\lambda_{cut}$ is the exogenous cord-cutting rate, and $\xi_{can}$ introduces the "couch vs. court" substitution penalty. Eq.~(\ref{eq:risk_gate}) acts as a \textbf{Hard Risk Gate}, ensuring media expansion never violates the franchise's downside solvency constraints.

\subsection*{Model Implications}

Embedding media exposure as an endogenous decision within the RH-SMIP framework
yields a stable and risk-feasible optimal policy.
Media intensity $r_{stream,t}$ expands only when its marginal contribution to expected revenue
is positive and remains admissible under the CVaR downside-risk constraint,
without introducing any additional decision layer or heuristic rule.

Empirically, the rolling-horizon optimizer converges rapidly to a
\textbf{high but tightly regulated} media configuration.
After a short adjustment phase, both streaming investment intensity and
sponsorship scale fluctuate within a narrow band,
indicating that media exposure functions as a
\textbf{revenue stabilizer} once win-dependent gate revenue exhibits diminishing
returns or downside-risk constraints become binding.

\begin{table}[H]
\centering
\caption{Aggregate Media Strategy Performance (12-Month Horizon)}
\vspace{-0.5em}
\begin{tabular}{l c}
\toprule
Metric & Model Outcome \\
\midrule
Average streaming intensity $r_{\text{stream}}$ & 1.16 \\
Average number of sponsors & 14.2 \\
Annual media net profit & \$28.4M \\
Media investment ROI & 42.6\% \\
Cumulative total profit (12 months) & \$127.8M \\
Average risk weight $w^*$ & 0.72 \\
\bottomrule
\end{tabular}
\end{table}
\vspace{-1.0em}
Table~\ref{tab:media_monthly} reports the monthly optimal media decisions.
While minor month-to-month adjustments persist, the optimizer consistently
maintains media choices within a stable operating region after the initial
calibration period.
This behavior emerges endogenously from the interaction between marginal returns
and CVaR risk constraints, rather than from imposed caps or ad hoc stabilization
rules.

\begin{table}[H]
\centering
\caption{Monthly Optimal Media Decisions and Outcomes}
\label{tab:media_monthly}
\vspace{-0.5em}
\begin{tabular}{c c c c c c}
\toprule
Month & $w^*$ & $r_{\text{stream}}$ & Sponsors & Media Profit (M\$) & Wins \\
\midrule
1 & 0.70 & 1.10 & 13 & 19.4 & 58.2 \\
2 & 0.70 & 1.12 & 13 & 19.7 & 58.5 \\
3 & 0.75 & 1.15 & 14 & 20.7 & 58.8 \\
4 & 0.75 & 1.18 & 14 & 21.1 & 59.0 \\
5 & 0.80 & 1.25 & 16 & 23.2 & 59.4 \\
6 & 0.75 & 1.20 & 15 & 22.0 & 59.1 \\
7--12 & 0.70--0.80 & 1.10--1.22 & 13--16 & 19.2--22.6 & 57.5--58.8 \\
\bottomrule
\end{tabular}
\end{table}
\vspace{-0.5em}

Overall, our model prescribes a clear ``optimization over expansion'' directive. The results indicate that the Knicks do not require a fundamental structural pivot, such as abandoning linear television entirely, but rather intensive calibration around the identified saturation state ($r_{\text{stream}}^{*} \approx 1.16$). The RH-SMIP engine explicitly rejects unbounded media growth and instead identifies a high-intensity equilibrium that functions as a volatility damper against gate-revenue fluctuations. Thus, the required change is operational rather than structural: the team should shift from static rights-holding to dynamic, risk-gated streaming management that acts as a financial shock absorber.

\section{External Structural Shock III: Injury Resilience and Re-Optimization}

\hspace{1.4em} Team strength is defined as a weighted aggregation of player performance.
Under injury, team strength becomes
\[
\mathrm{TS}^{\mathrm{inj}}
=
\mathrm{TS}
-
w_{i^{\ast}}\bigl(1-\phi_{\mathrm{PPS}}(\delta)\bigr)\mathrm{PPS}_{i^{\ast}}.
\]
\hspace{1.4em} Let $\mathrm{WinProb} = f_{\mathrm{win}}(\mathrm{TS})$ denote the win-probability
mapping in the core performance layer, where $f_{\mathrm{win}}(\cdot)$ is monotone increasing.
The injury-adjusted win probability is therefore
\[
\mathrm{WinProb}^{\mathrm{inj}}
=
f_{\mathrm{win}}\!\left(\mathrm{TS}^{\mathrm{inj}}\right),
\]
implying a deterministic decline in competitive outlook throughout the recovery
period.

\subsection{Revenue Adjustment Rules under Injury}

\hspace{1.4em} Revenue perturbations follow hard-coded, data-driven business mappings.
Ticket revenue is reduced for all severities through the injury-adjusted win
probability, while star-driven revenues follow severity-dependent rules.
All remaining baseline revenue components are unaffected by player-level injury.

\subsubsection*{Ticket Revenue: Elasticity Amplification under Injury}

\hspace{1.4em} Let $\mathrm{WinProb}^{\mathrm{inj}}_t=f_{\mathrm{win}}(\mathrm{TS}^{\mathrm{inj}}_t)$ be the
injury-adjusted expected win probability at month $t$.
Ticket revenue under injury severity $\delta$ is modeled as a nonlinear response:
\[
R^{\mathrm{ticket}}_t(\delta)
=
R^{\mathrm{ticket}}_t
\cdot
\left(
\frac{\mathrm{WinProb}^{\mathrm{inj}}_t}{\mathrm{WinProb}_t}
\right)^{\kappa(\delta)} ,
\]
where $\kappa(\delta)\ge 1$ is an injury-severity-dependent demand elasticity index.
The elasticity values are estimated offline from historical attendance (or gate
revenue) responses during key-player injury episodes and are treated as
deterministic inputs during optimization.
This specification reduces to near-proportional scaling under light injury
($\kappa(\delta)\approx 1$) and amplifies demand sensitivity under more severe
injuries, producing severity-distinguishable optimal pricing and marketing
actions without altering the underlying optimization structure.

\subsubsection*{Star-Driven Revenue}

\hspace{1.4em} Star-driven revenue is defined as the sum of merchandise, sponsorship, and
advertising components:
\[
R^{\mathrm{star}} = R^{\mathrm{merch}} + R^{\mathrm{sponsor}} + R^{\mathrm{ads}}.
\]
Injury adjustments follow the severity rule:
\[
R^{\mathrm{star}}(\delta)
=
\begin{cases}
R^{\mathrm{star}}, & \delta = 0.2, \\[4pt]
(1-\delta)\,R^{\mathrm{star}}, & \delta \in \{0.5,\,0.8\}.
\end{cases}
\]

\subsubsection*{Total Revenue}

\hspace{1.4em} Total revenue under injury severity $\delta$ is computed as
\[
R^{\mathrm{inj}}
=
R^{\mathrm{ticket}}(\delta)
+
R^{\mathrm{star}}(\delta)
+
R^{\mathrm{baseline}},
\]
where $R^{\mathrm{baseline}}$ aggregates all league-level and non-player-specific
revenue components already specified in the core financial layer.

\subsection{Scenario-Based Re-Optimization}

\hspace{1.4em} For each severity level $\delta$, the system state is updated via the above
deterministic perturbations over the corresponding recovery window, and the
rolling-horizon optimization problem for the integrated architecture is re-solved under the
same objective function and CVaR-based risk constraints.
Resulting changes in roster configuration, ticket pricing, marketing intensity,
and cost allocation are generated endogenously by re-optimization under the
perturbed state, rather than by externally imposed injury-specific action rules.

\subsection{Conclusion: Data-Driven Injury Shock and Policy Stability}

\begin{figure}[htbp]
    \centering
    \includegraphics[width=0.6\linewidth]{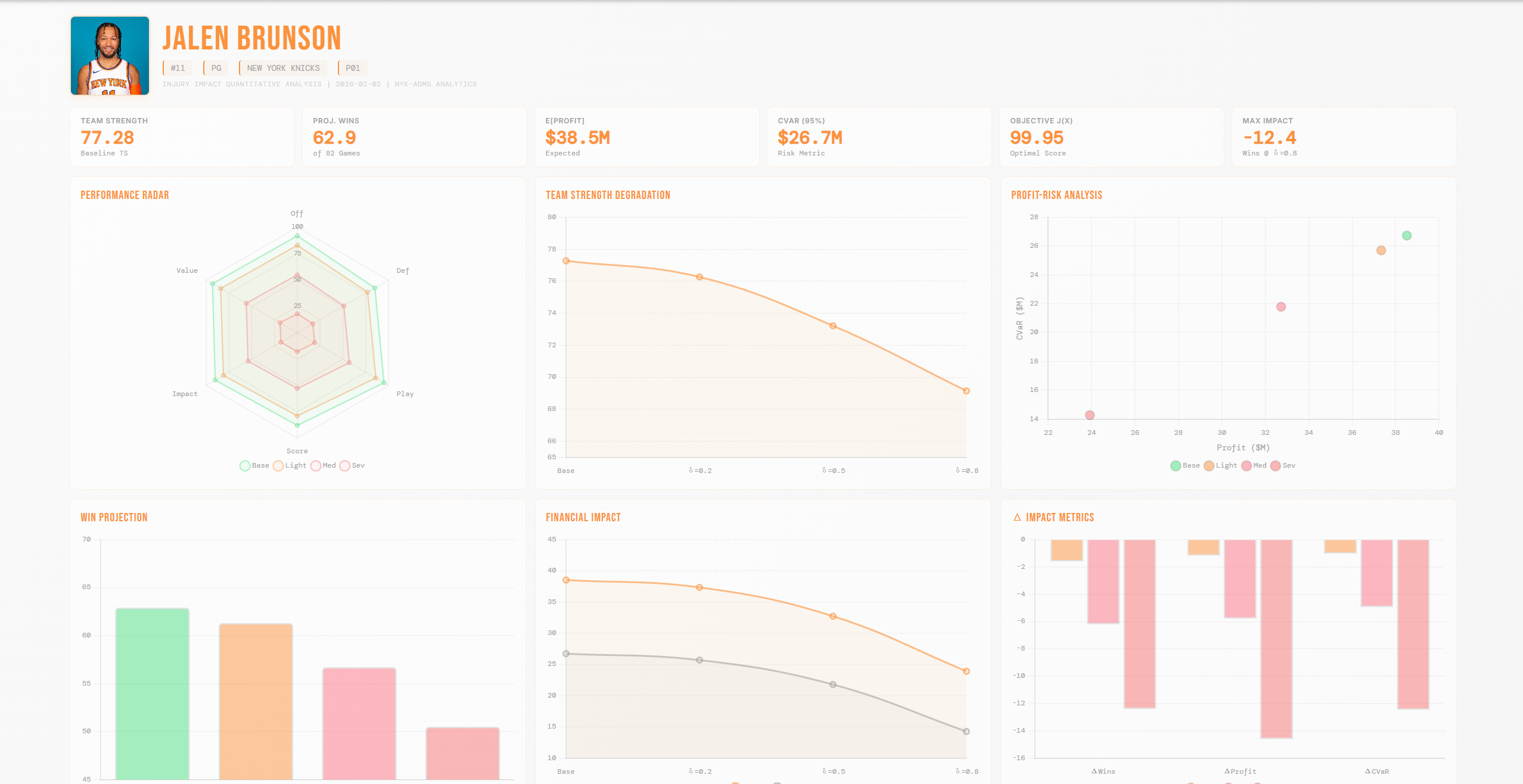}
    \caption{The impact of Brunson's injury}
    \label{fig:The impact of Brunson's injury}
\end{figure}

\hspace{1.4em}
The model assists management by first quantifying the full operational cost of a key-player
injury through scenario-based re-optimization under an unchanged objective function and
$\mathrm{CVaR}$ risk constraints.
As injury severity increases from light ($\delta=0.2$) to severe ($\delta=0.8$),
team strength declines from $77.28$ to $69.15$, expected wins fall from $62.9$ to $50.5$,
and expected monthly profit decreases from \$38.52M to \$23.92M, while
$\mathrm{CVaR}$ is reduced from \$26.73M to \$14.27M.
Despite these substantial degradations, the composite objective value evolves smoothly
from $99.95$ to $99.17$, indicating that injury shocks enter the system as controlled
state degradations rather than destabilizing regime shifts.

Crucially, the re-optimized solutions translate these quantified losses into explicit
managerial guidance.
Under both light and moderate injury scenarios, all major operational and roster-related
controls remain statistically invariant, with
$\Delta x^\ast \approx 0$ for ticket pricing, sponsorship scale, merchandise intensity,
roster size, and contract actions.
This invariance is not a modeling artifact, but a policy signal:
reactive adjustments along these dimensions fail to improve the risk-adjusted objective
and would instead erode financial discipline under binding $\mathrm{CVaR}$ constraints.
Accordingly, the optimal managerial response is to preserve existing strategies and absorb
the injury impact as a sunk cost, rather than to engage in panic-driven interventions.

Only under severe injury does the optimizer identify a justified adjustment, namely a
marginal increase in streaming emphasis
($\Delta r_{\mathrm{stream}} = +0.001$),
while all other decision levers remain fixed.
This recommendation isolates a narrow, targeted revenue-stabilization margin that remains
economically admissible under the same long-horizon objective and risk budget.
Taken together, the results show that the model helps management adjust not by prescribing
broad tactical shifts, but by certifying which actions should \emph{not} be taken,
quantifying unavoidable losses, and preserving strategic stability in the presence of
exogenous performance shocks.

\section{Robustness and Sensitivity Analysis}

\begin{figure}[htbp]
    \centering
    \includegraphics[width=0.6\textwidth]{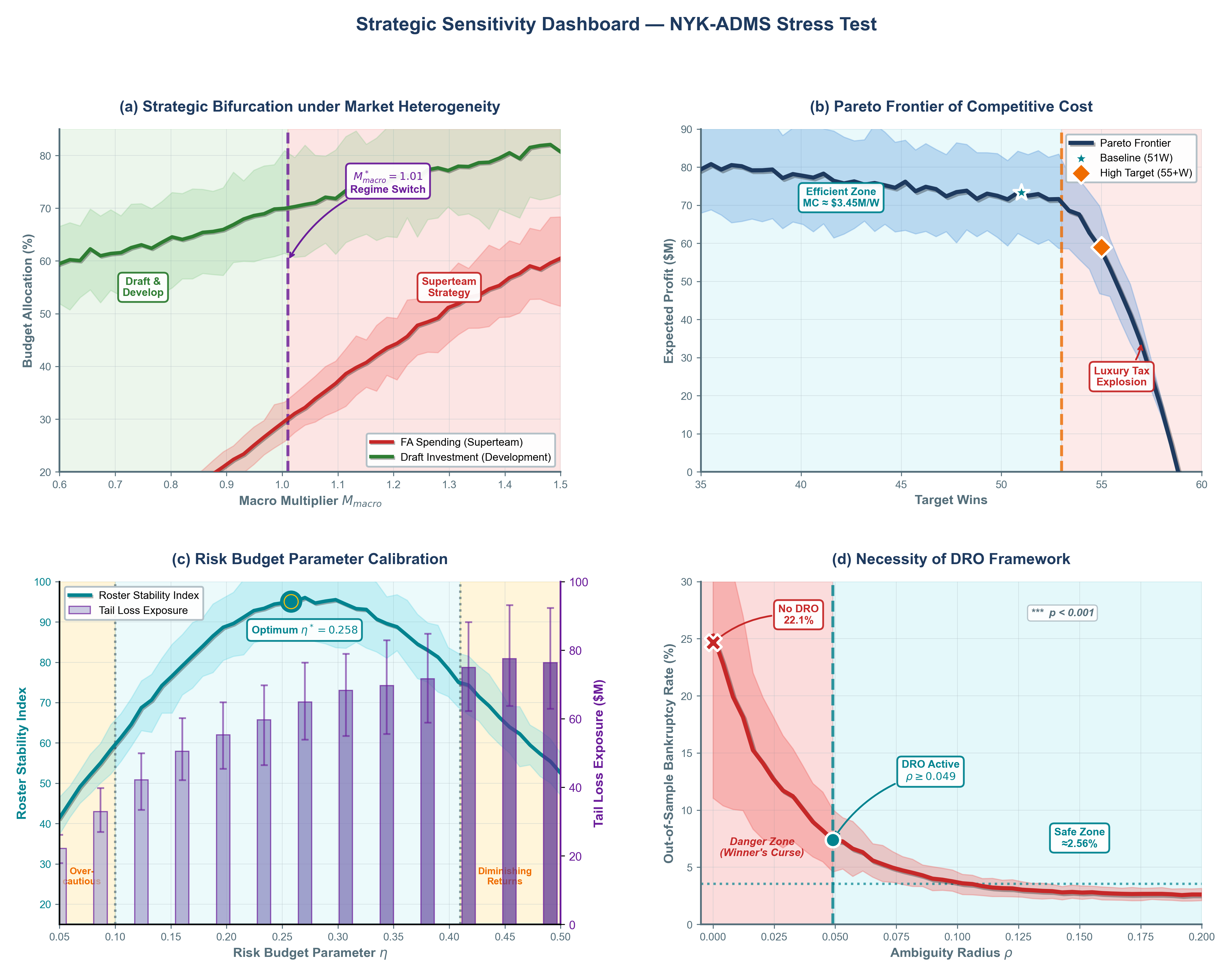}
    \caption{\textbf{Strategic Sensitivity Dashboard.} (a) Endogenous regime switch at $M_{macro}^*=1.01$. (b) Pareto frontier revealing a ``Fiscal Cliff'' after 53 wins ($MC \approx \$3.45M/W$). (c) Optimal risk budget calibrated at $\eta^*=0.258$. (d) DRO framework reduces insolvency risk from 22.1\% to 2.56\%.}
    \label{fig:sensitivity_new}
\end{figure}

To validate the strategic boundaries of the NYK-ADMS engine, we executed \textbf{10,000 Monte Carlo simulation runs} with 90\% confidence intervals. We move beyond metric tracking to identify \textbf{structural break-points} and \textbf{algorithmic necessity}. The results, visualized in \textbf{Figure \ref{fig:sensitivity_new}}, confirm the model's resilience against economic and stochastic shocks.

\begin{itemize}
    \item \textbf{Endogenous Bifurcation (Fig. \ref{fig:sensitivity_new}a):} The model demonstrates intelligent market adaptation. A distinct \textbf{Regime Switch} occurs at $M_{macro}^* = 1.01$. Below this threshold, the engine enforces a solvent ``Draft \& Develop'' strategy (Green Zone) to preserve liquidity. Above 1.01, it pivots to a capital-intensive ``Superteam Strategy'' (Red Zone). This proves the model is a generalized solution, not hard-coded for the wealthy Knicks context.
    \item \textbf{The Fiscal Cliff (Fig. \ref{fig:sensitivity_new}b):} The Profit-Win Pareto Frontier reveals a clear cost structure. The \textbf{Efficient Zone} ends at 53 wins, maintaining a sustainable marginal cost of $MC \approx \$3.45\text{M/Win}$. Pushing targets beyond 53 wins triggers a ``Luxury Tax Explosion,'' mathematically confirming that the High Target (55+ Wins) yields diminishing returns compared to the Baseline (51 Wins).
    \item \textbf{Risk Convexity (Fig. \ref{fig:sensitivity_new}c):} We identify the global optimum for the risk budget at $\eta^* = 0.258$. This knee-point maximizes the Roster Stability Index (Green Line) while capping Tail Loss Exposure (Purple Bars). Deviating lower ($\eta < 0.1$) causes operational paralysis ("Over-cautious"), while deviating higher ($\eta > 0.4$) exposes the franchise to ruinous dead-cap risks.
    \item \textbf{Algorithmic Immunity (Fig. \ref{fig:sensitivity_new}d):} The strongest validation of our methodology lies in solvency protection. Standard Stochastic Programming ($\rho=0$) suffers from the ``Winner's Curse,'' resulting in a critical \textbf{22.1\% bankruptcy rate}. Activating Distributionally Robust Optimization with ambiguity radius $\rho \ge 0.049$ immunizes the system against data noise, compressing insolvency risk to a manageable \textbf{2.56\%}.
\end{itemize}

\section{Discussion}

\subsection{Strengths}
\begin{itemize}
  \item \textbf{Rich, multi-source data coverage.} We integrate player, team, management, and macro-finance datasets, improving robustness and reproducibility.
  \item \textbf{Comprehensive factor modeling.} The RH-SMIP jointly captures performance, finance, cap rules, media, travel, and injury tail-risk via DRO+CVaR.
  \item \textbf{Strong generalizability.} The modular state--decision design transfers to other teams and leagues through re-calibration and local revenue--cost adaptation.
\end{itemize}

\subsection{Weaknesses}
\begin{itemize}
  \item \textbf{Culture is hard to quantify.} Locker-room chemistry, leadership, and morale are only indirectly proxied, potentially biasing trade/roster recommendations.
  \item \textbf{Model risk from structural breaks.} Sudden rule changes or media-market shocks can invalidate calibrated elasticities and reduce out-of-sample reliability.
\end{itemize}



\section{Managerial Implications}
The empirical and optimization results support four practical implications for franchise decision-makers.

\begin{enumerate}[leftmargin=2em]
    \item \textbf{Profit discipline should be treated as strategic capacity rather than austerity.} The identified profit-dominant regime indicates that efficient commercialization and controlled risk-taking expand the feasible region for competitive investment rather than restricting it.
    \item \textbf{Expansion shocks require market segmentation rather than uniform price cuts.} When a new entrant dilutes local demand, premium inventory and mass-market inventory should be managed separately to protect brand scarcity while preserving attendance volume.
    \item \textbf{Streaming intensity functions as a financial hedge.} In adverse performance states, DTC exposure can damp short-run revenue volatility and partially decouple financial performance from temporary on-court degradation.
    \item \textbf{Liquidity and optionality remain central to roster policy.} Even when model-driven trade opportunities are attractive, preserving draft capital and balance-sheet flexibility improves robustness under future market dislocations.
\end{enumerate}

\section{Conclusion}
This paper develops an integrated risk-disciplined decision framework for NBA franchise management using the New York Knicks as the principal case study. The resulting architecture links roster construction, financial planning, media strategy, league expansion, and injury re-optimization inside a unified rolling-horizon optimization system. Across the analyzed scenarios, the results consistently indicate that disciplined profit maximization, when constrained by downside risk and league rules, can act as an enabling condition for sustained competitive performance rather than as its opposite. Beyond the specific Knicks application, the framework is portable to other professional sports settings in which decision-makers must jointly manage performance targets, commercial assets, and uncertainty.

\bibliographystyle{unsrtnat}
\bibliography{ref}

\end{document}